\documentclass[sn-mathphys,Numbered]{sn-jnl}

\usepackage{graphicx}%
\usepackage{multirow,multicol}%
\usepackage{amsmath,amssymb,amsfonts}%
\usepackage{amsthm}%
\usepackage{mathrsfs}%
\usepackage[title]{appendix}%
\usepackage{xcolor}%
\usepackage{textcomp}%
\usepackage{manyfoot}%
\usepackage{booktabs}%
\usepackage{algorithm}%
\usepackage{algorithmicx}%
\usepackage{algpseudocode}%
\usepackage{listings}%

\usepackage{soul}
\usepackage{chemformula}
\usepackage[version=3]{mhchem} 
\usepackage{physics}
\usepackage{verbatim}

\theoremstyle{thmstyleone}%
\theoremstyle{thmstyletwo}%

\theoremstyle{thmstylethree}%

\theoremstyle{definition}
\newtheorem{thm}{Theorem}
\newtheorem{dfn}[thm]{Definition}

\newcommand{\R}{\mathbb{R}}
\newcommand{\Z}{\mathbb{Z}}

\newcommand{\CRIS}{\mathrm{CRIS}}
\newcommand{\PDD}{\mathrm{PDD}}

\newcommand{\AMD}{\mathrm{AMD}}
\newcommand{\ADA}{\mathrm{ADA}}
\newcommand{\PDA}{\mathrm{PDA}}
\newcommand{\LND}{\mathrm{LND}}

\newcommand{\EMD}{\mathrm{EMD}}

\newcommand{\PPC}{\mathrm{PPC}}

\newcommand{\ep}{\varepsilon}

\newcommand{\vol}{\mathrm{vol}}

\newcommand{\angstrom}{\textup{\AA}}

\graphicspath{{images/}}

\begin{document}

\title[Geographic-style maps of the materials space]{Geographic-style maps with a local novelty distance help navigate the materials space}


\author[1]{\fnm{Daniel} \sur{Widdowson}}

\author*[1]{\fnm{Vitaliy} \sur{Kurlin}}\email{vitaliy.kurlin@liverpool.ac.uk}

\affil[1]{\orgdiv{Materials Innovation Factory}, \orgname{University of Liverpool}, \orgaddress{\street{Oxford Street}, \city{Liverpool}, \postcode{L7 3NY}, \country{United Kingdom}}}



\abstract{
With the advent of self-driving labs promising to synthesize large numbers of new materials, new automated tools are required for checking potential duplicates in existing structural databases before a material can be claimed as novel.
To avoid duplication, we rigorously define the novelty metric of any periodic material as the smallest distance to its nearest neighbor among already known materials.
\smallskip

Using ultra-fast structural invariants, all such nearest neighbors can be found within seconds on a typical computer even if a given crystal is disguised by changing a unit cell, perturbing atoms, or replacing chemical elements.
This real-time novelty check is demonstrated by finding near-duplicates of the 43 materials produced by Berkeley's A-lab in the world's largest collections of inorganic structures, the Inorganic Crystal Structure Database and the Materials Project.
\smallskip

To help future self-driving labs successfully identify novel materials, we propose navigation maps of the materials space where any new structure can be quickly located by its invariant descriptors similar to a geographic location on Earth.
}

\keywords{materials space, crystal structure, isometry invariant, continuous metric}



\maketitle

\vspace*{-6mm}
\section{Introduction: how is the materials space defined?}
\label{sec:intro}

The chemical space of all possible molecules is often estimated at the scale of $10^{60}$ \cite{orsi2025navigating}.
Similar numbers are quoted for potential materials, though {many} polymorphs such as diamond and graphite have the same chemical composition and hence can only be distinguished by their geometry.
When materials are claimed to be novel amongst already known ones, we need to rigorously define what constitutes two materials being the ``same or different'' \cite{sacchi2020same}.
The definition of a \emph{crystal structure} was finalized in the periodic case in \cite{anosova2024importance}, so we focus on ideal periodic crystals (briefly,  \emph{crystals}) as formalized below.   
When a material is disordered, we consider its closest periodic analogue. 
\smallskip

\begin{figure}[h!]
\caption{Almost any tiny perturbation discontinuously scales up a primitive cell and makes unreliable any comparison based on cells or motifs. This discontinuity was resolved without relying on cells \cite{widdowson2022resolving,widdowson2025higher}.}
\includegraphics[width=\textwidth]{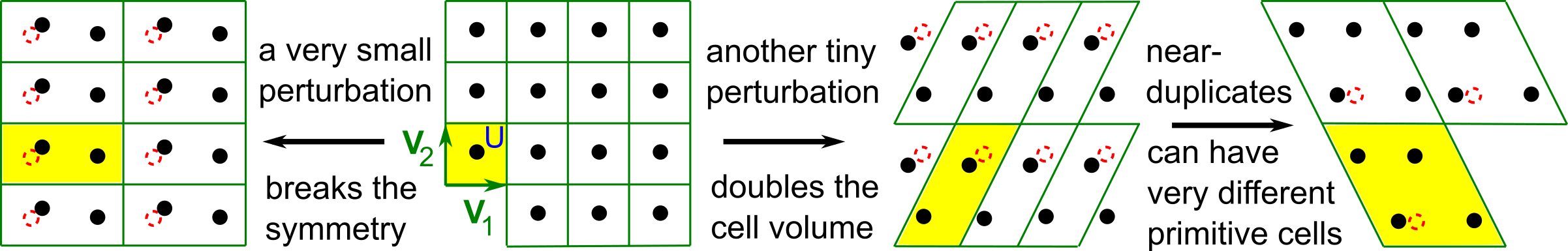}
\label{fig:perturbations}
\end{figure}

A \emph{crystal} is usually given by a basis of vectors $\vb*{v_1},\vb*{v_2},\vb*{v_3}$ in Euclidean space $\R^3$ and a \emph{motif} of atoms with chemical elements and fractional coordinates in this basis.
If we forget about chemical elements, the atomic centers $p_1,\dots,p_m$ can be considered zero-sized points in the \emph{primitive unit cell} $U=\{t_1 \vb*{v_1}+t_2\vb*{v_2}+t_3\vb*{v_3} \mid t_1,t_2,t_3\in[0,1)\}$ defined by the basis $\vb*{v_1},\vb*{v_2},\vb*{v_3}$.
In dimension 2, the second picture of Fig.~\ref{fig:perturbations} highlights the square cell $U$ with the orthonormal basis $\vb*{v_1},\vb*{v_2}$.
Then the underlying \emph{periodic point set} of any crystal consists of infinitely many points $p_i+c_1\vb*{v_1}+c_2\vb*{v_2}+c_3\vb*{v_3}$ for $i=1,\dots,m$ and integer coefficients $c_1,c_2,c_3\in\Z$. 
Infinitely many different pairs of a basis (or a primitive cell) and a motif $M$ generate pointwise identical crystals, see a detailed discussion of this ambiguity of the traditional definition in \cite[section~2]{anosova2024importance}.

\vspace*{-2mm}

\vspace*{-2mm}
Since atoms always vibrate \cite[chapter~1]{feynman1971feynman}, their fractional coordinates are always uncertain and will slightly deviate under repeated measurements even on the same instrument.
Almost any displacement of one atom breaks the symmetry and can arbitrarily scale up a primitive unit cell as in Fig.~\ref{fig:perturbations}.
This discontinuity of a reduced cell \cite{niggli1928krystallographische} was experimentally reported in 1965 \cite[p.~80]{lawton1965reduced} and remained unresolved until 2022 \cite{widdowson2022resolving} when all periodic crystals in the Cambridge Structural Database (CSD) \cite{ward2020introduction} were distinguished within two days (now within an hour) on a modest desktop computer.
Several unexpected duplicates with identical geometries (almost to the last decimal place in all cell parameters and atomic coordinates) but with different chemistry are under investigation by five journals for data integrity \cite[section~6]{widdowson2022average}. 
\smallskip

Since crystal structures are determined in a rigid form, there is no sense in distinguishing crystal representations related by a \emph{rigid motion} (a composition of translations and rotations in $\R^3$), which change a basis and atomic coordinates.
On the other hand, there is no sense to fix any threshold $\ep>0$ that would allow us to call crystals the ``same'' if all their atomic centers (without chemical attributes) can be matched up to $\ep$-perturbations.
Indeed, any periodic point sets can be connected by sufficiently many $\ep$-perturbations \cite[Proposition~2.10]{widdowson2022average}, which makes the classification based on any threshold $\ep>0$ trivial due to the transitivity axiom saying that if $S$ is equivalent to $Q$, and $Q$ is equivalent to $T$, then $S$ is equivalent to $T$ \cite[section~1]{anosova2024importance}. 
\smallskip

Hence a rigorous way to classify crystals under rigid motion, is to define the \emph{crystal structure} as a rigid class of periodic point sets, see \cite[Definition~6]{anosova2024importance}.
Then any deviations of atomic positions are not ignored but continuously quantified by a distance metric between different rigid classes.
This definition would remain impractical unless we can efficiently separate rigid classes by quickly computable \emph{invariants} that are numerical properties preserved under rigid motion.
The chemical composition written as percentages of chemical elements is such an invariant but is \emph{incomplete} because many polymorphs have the same composition but can not be matched by rigid motion.
\smallskip

In the sequel, we will consider the sightly weaker equivalence of \emph{isometry} (any distance-preserving transformation in $\R^3$), which is a composition of rigid motion and {mirror} reflections.
Since mirror images can be distinguished by a suitable sign of orientation, the main difficulty is to classify periodic point sets under isometry.
\smallskip

When comparing crystals as periodic sets of atomic centers without chemical attributes, it might seem that all chemistry is lost.
However, the fact that all (more than 850 thousand) periodic crystals in the CSD (apart from the investigated duplicates) can be distinguished by isometry invariants in section~\ref{sec:methods} implies that no information is lost so that all chemistry under standard conditions such as temperature and pressure is in principle reconstructable from sufficiently precise atomic geometry. 
\smallskip

\begin{figure}[h!]
\caption{\textbf{Left}: the \emph{Crystal Isometry Principle} says that all chemistry of any real periodic crystal under standard ambient conditions can be reconstructed from (the isometry class of) the periodic set of atomic centers given with precisely enough coordinates \cite{widdowson2022resolving}.
\textbf{Right}: most optimization methods output local optima without exploring the space around. De-fogging this \emph{Crystal Isometry Space} $\CRIS(\R^3)$ beyond known or predicted materials will enable a proper navigation across the crystal universe.}
\smallskip

\includegraphics[height=26mm]{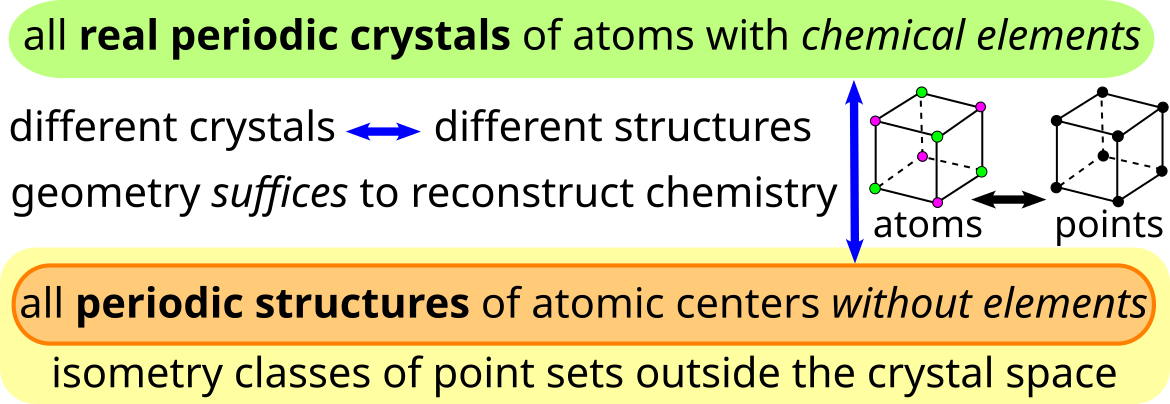}
\hspace*{1mm}
\includegraphics[height=26mm]{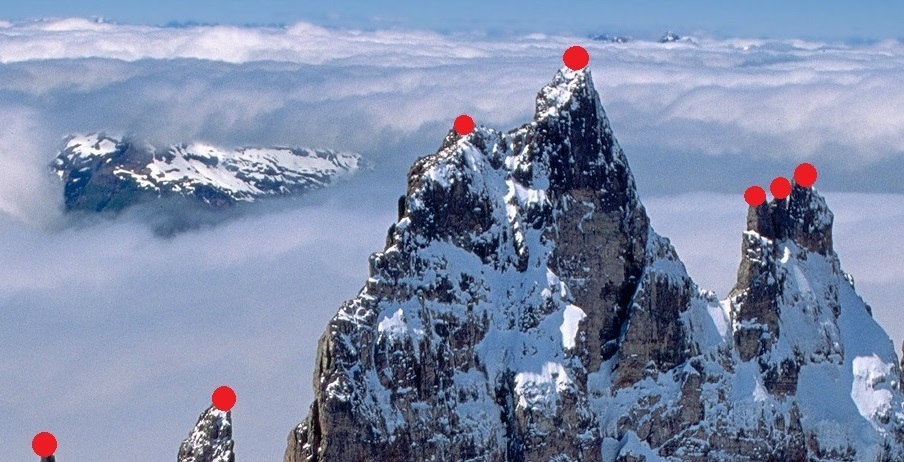}
\label{fig:CRISP}
\end{figure}

This Crystal Isometry Principle (CRISP) first appeared in 2022 \cite[section~7]{widdowson2022average} and was inspired by Richard Feynman's hint in   \cite[chapter~1, Fig.1-7]{feynman1971feynman}, which distinguished 7 cubic crystals by their cube size in the first lecture ``Atoms {in} motion'', see Fig.~\ref{fig:CRISP}~(left).
\smallskip

More importantly, when we 
consider atoms only as zero-sized points, we can study all periodic structures in a common space {similar to the periodic table of all elements}.
\smallskip

{In the geographic analogy, the chemical composition can be compared to the altitude (the height above the sea level) of any location on Earth. 
If our geographic map is precise enough, we can determine the average temperature or any other property at every location.
If we know the altitude (chemical composition) in addition to geographic coordinates (structural invariants), the property prediction will be easier.}
\medskip

\begin{dfn}[space of periodic materials]
\label{dfn:CRIS}
The \emph{Crystal Isometry Space} $\CRIS(\R^3)$ is the space of isometry classes of all periodic sets of points without atomic attributes.
\end{dfn}
\smallskip

Since (the isometry class of) any periodic point set has a unique location in $\CRIS(\R^3)$, all known materials can be considered `visible stars' in this continuous universe.
Any periodic crystal discovered in the future will appear at a new unique {location like a `new star', while all past crystals remain at the same locations.
Until recently, optimizing complicated energy functions blindly climbed as a mountaineer to a high peak, illustrated in Fig.~{\ref{fig:CRISP}}~(right), and stopping at many (approximations to) isolated peaks without even any reliable method to continuously measure distances between these peaks, while the remaining landscape was covered by clouds. 
Our vision is to map the full space $\CRIS(\R^3)$ 
to enable a non-blind discovery of materials} \cite{anosova2021introduction,anosova2026geometric}.
\smallskip

If we do not restrict the motif size, the space $\CRIS$ is infinitely dimensional.
However, if we consider all periodic sets with $m$ points in a motif, the resulting subspace $\CRIS(\R^3;m)$ has dimension $3m+3$ due to $m$ triples $x,y,z$ of atomic coordinates and $6$ parameters of a unit cell, of which 3 are neutralized by translations along basis vectors.
Alternatively, we can define a unit cell by 3 basis vectors with 3 coordinates, of which 6 are neutralized by 3+3 parameters of translations and rotations in $\R^3$.  
\smallskip

In the partial case $m=1$, $\CRIS(\R^3;1)$ is a continuous 6-dimensional space of 3D lattices, which was previously cut in 14 disjoint subspaces of Bravais classes \cite{bravais1850memoir} but is now parametrized by complete invariants \cite{kurlin2022complete,bright2021welcome}.
Continuous maps of the simpler 3-dimensional space $\CRIS(\R^2;1)$ of 2D lattices recently appeared in \cite{kurlin2024mathematics}, \cite{bright2023geographic}, \cite{bright2023continuous}.
\smallskip

The full space $\CRIS(\R^3)=\cup_{m=1}^{+\infty} \CRIS(\R^3;m)$ is a union of infinitely many subspaces for $m=1,2,3,\dots$ such that any periodic set with $m$ points in a cell is infinitesimally close to infinitely many subspaces of sets with $2m,3m,\dots$ points in a primitive cell.
Indeed, perturbations in Fig.~\ref{fig:perturbations} arbitrarily extend any given cell and make the extended cell primitive by a tiny displacement of any atom and all its translational copies. 
Crystals should be continuously compared only across multiple subspaces, not within one subspace $\CRIS(\R^3;m)$ for a fixed number $m$ of atoms. 
Any database of periodic crystals is a finite sample from the continuous space $\CRIS(\R^3)$.
\smallskip

The {first contribution of this work} is the local novelty distance based on generically complete invariants, {which identify} closest neighbors of the 43 A-lab crystals in the Inorganic Crystal Structure Database (ICSD) \cite{zagorac2019recent} and Materials Project (MP) \cite{jain2013commentary} {within seconds on a desktop computer.
The second contribution is the geographic-style maps showing how the ICSD and MP populate $\CRIS(\R^3)$ in invariant coordinates}.
 
\section{Methods: invariant-based novelty distance metric}
\label{sec:methods}

This section introduces a new metric LND (Local Novelty Distance) that satisfies all metric axioms and continuously quantifies in real time a deviation of any newly synthesized crystal from its nearest neighbor in an existing structural database.  

\subsection{Generically complete and continuous structural invariants}
\label{sub:invariants}
 
Definition~\ref{dfn:PDD} reminds us of the Pointwise Distance Distribution (PDD), which suffices together with a lattice to reconstruct any generic periodic point set $S\subset\R^3$ under isometry by \cite[Theorem 4.4]{widdowson2022resolving} and \cite[Theorem~5.8]{widdowson2026pointwise}.
Generic means any set apart from a singular subspace of measure 0, e.g. almost any noise makes every crystal generic.
\smallskip

The PDD is a matrix of inter-point distances and is stronger than the Pair Distribution Function (PDF) \cite{terban2022structural} in the sense that PDD can be simplified to PDF  but distinguishes homometric structures \cite{patterson1939homometric} that have the same PDF \cite[section~3]{widdowson2022resolving}. 
\smallskip

\begin{dfn}[isometry invariant $\PDD(S;k)$]
\label{dfn:PDD}
Let $S\subset\R^n$ be a periodic point set with a motif $M=\{p_1,\dots,p_m\}$.
Fix an integer $k\geq 1$.
For every point $p_i\in M$, let $d_1(p)\leq\dots\leq d_k(p)$ be the distances from $p$ to its $k$ nearest neighbors within the full infinite set $S$ not restricted to any cell.
The matrix $D(S;k)$ has $m$ rows consisting of the distances $d_1(p_i),\dots,d_k(p_i)$ for $i=1,\dots,m$.
If any $l\geq 1$ rows are identical to each other, we collapse them into a single row and assign the weight $l/m$ to this row.
The resulting matrix of maximum $m$ rows and $k+1$ columns including the extra (say, $0$-th) column of weights is called the \emph{Pointwise Distance Distribution} $\PDD(S;k)$. 
\end{dfn}
\smallskip

In Definition~\ref{dfn:PDD}, any point $p_i\in M$ can have several different neighbors at the same distance but the $k$ smallest distances (without any indices or types of neighbors) are always well-defined.
The matrix $\PDD(S;k)$ has ordered columns (according to the index of neighbors) but 
unordered rows because points of a motif of $S$ are unordered.
{The appendix computes a weighted PDD with atomic masses as extra weights of rows in $\PDD(S;k)$.
Using only on atomic centers detects} duplicates where chemical elements were artificially replaced without changing geometry, see \cite[Table~1]{anosova2024importance}.
\smallskip

If the number $k$ of neighbors increases to infinity, the asymptotic behavior of distances to neighbors is described in terms of the Point Packing Coefficient below. 
\smallskip

\begin{dfn}[Point Packing Coefficient $\PPC$]
\label{dfn:PPC}
Let $S\subset\R^3$ be a periodic point set with $m$ atoms in a unit cell $U$.
The \emph{Point Packing Coefficient} is $\PPC(S)=\sqrt[3]{\dfrac{\vol(U)}{mV_3}}$, where $\vol(U)$ is the volume of $U$, $V_3=\dfrac{4}{3}\pi$ is the volume of the unit ball in $\R^3$.
\end{dfn} 
\smallskip

The distances in each row of $\PDD(S;k)$ asymptotically increase as $\PPC(S)\sqrt[3]{k}$ by \cite[Theorem~13]{widdowson2022average}.
This asymptotic behavior motivates the simplified invariants below.
\smallskip
 
\begin{figure}[h!]
\caption{{The average invariants $\AMD_k$ and $\ADA_k$ from Definition}~\ref{dfn:PDA} {for $k=1,\dots,25$ and five simple crystals from the Materials Project, see more details and perovskite examples in the appendix.}}
\includegraphics[width=0.49\textwidth]{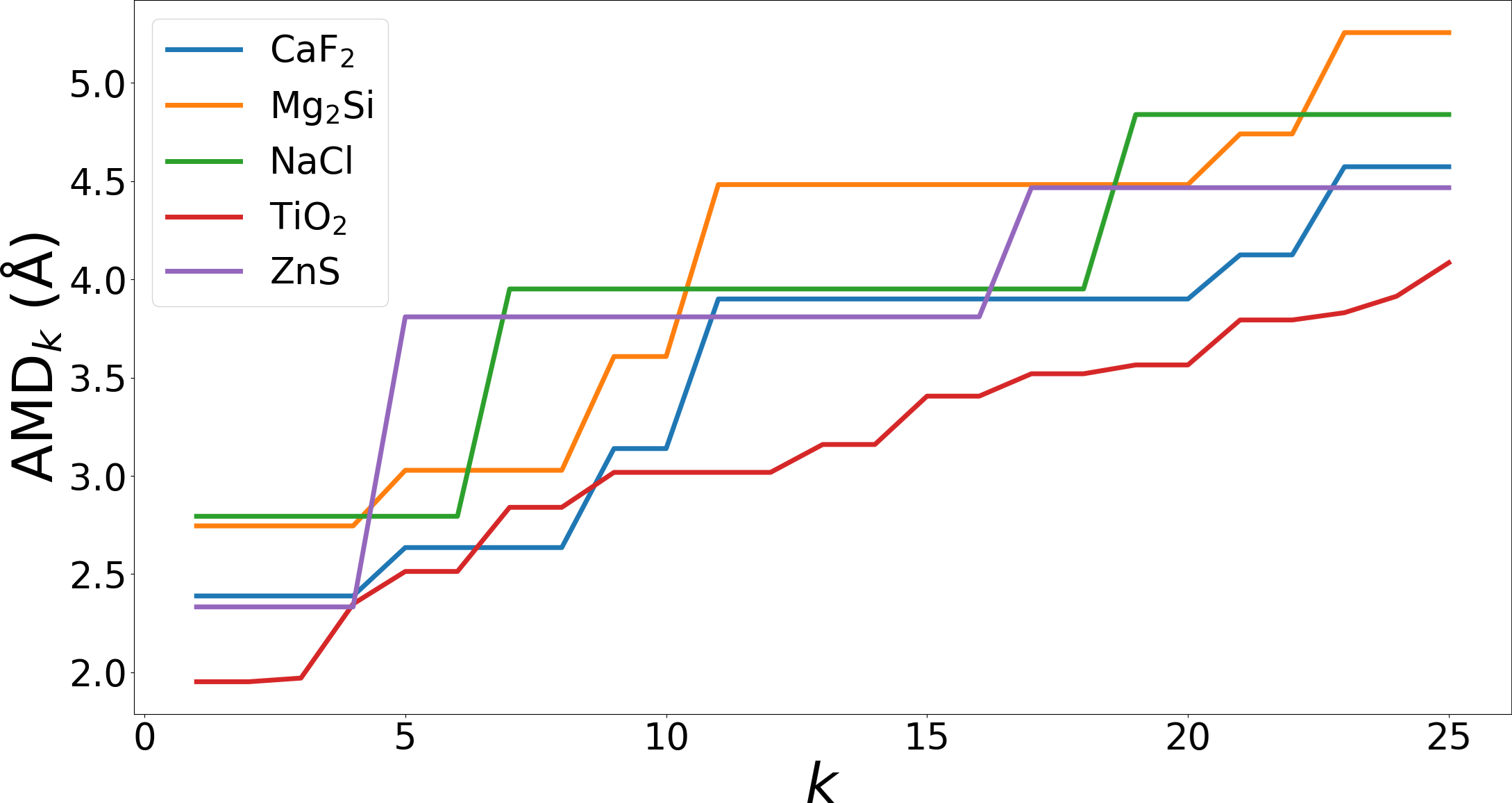}
\includegraphics[width=0.49\textwidth]{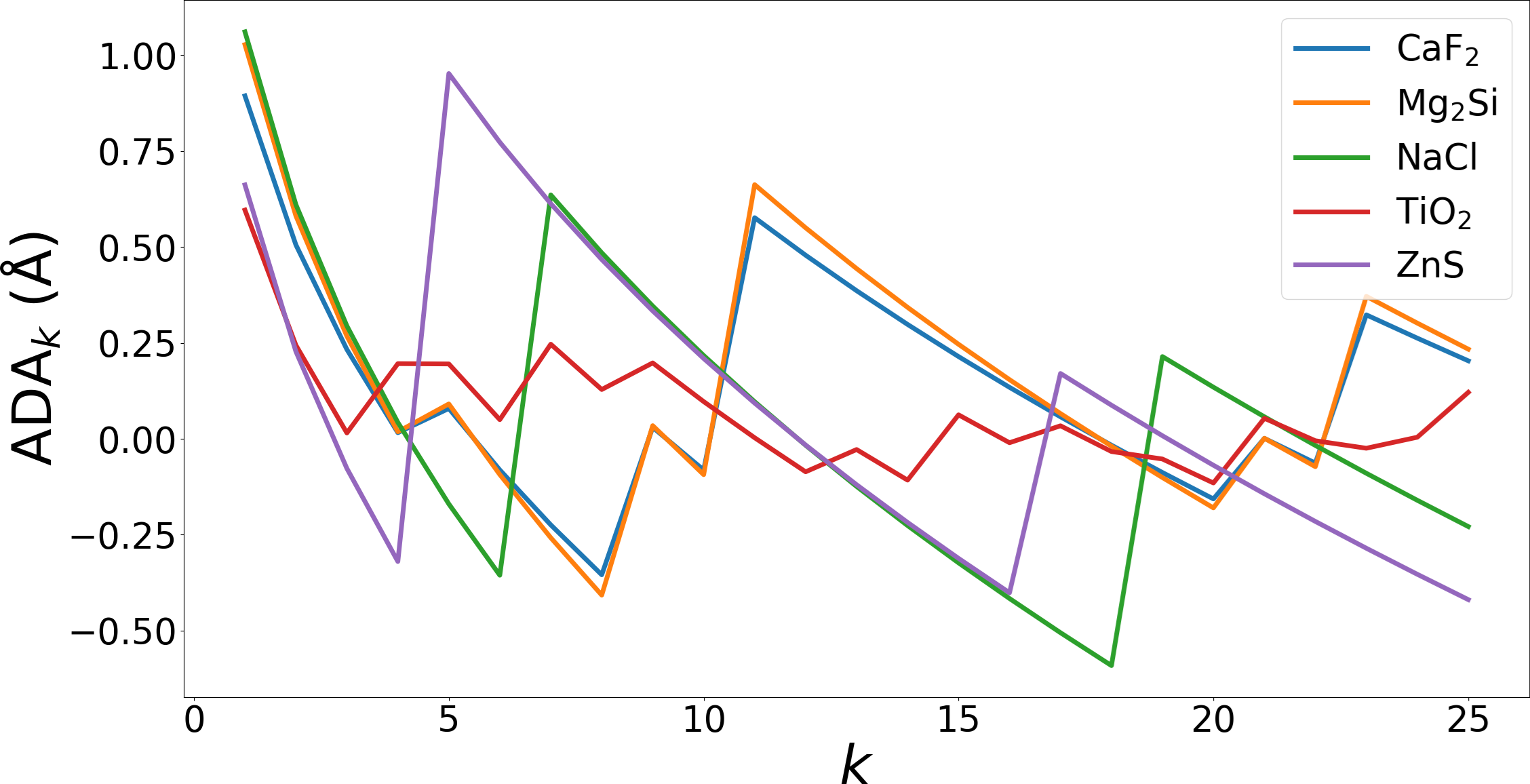}
\label{fig:AMD25ADA_5common_crystals}
\end{figure}

\begin{dfn}[invariants $\AMD$, $\ADA$, $\PDA$]
\label{dfn:PDA}
The \emph{Average Minimum Distance} $\AMD_k(S)$ is the weighted average of the $k$-th column of $\PDD(S;k)$. 
The \emph{Average Deviation from Asymptotic} is $\ADA_k(S)=\AMD_k(S)-\PPC(S)\sqrt[3]{k}$ for $k\geq 1$.
The \emph{Pointwise Deviation from Asymptotic} is the matrix $\PDA(S;k)$ obtained from $\PDD(S;k)$ by subtracting $\PPC(S)\sqrt[3]{k}$ from any distance in row $i$ and column $k$ for $i,k\geq 1$, see Fig.~\ref{fig:AMD25ADA_5common_crystals}.
\end{dfn} 

\vspace*{-2mm}
\vspace*{-2mm}

The invariants $\AMD_k$ and $\ADA_k$ form vectors of length $k$, e.g. set $\AMD(S;k)=(\AMD_1(S),\dots,\AMD_{k}(S))$ and $\ADA(S;k)=(\ADA_1(S),\dots,\ADA_{k}(S))$.
These vectors can be compared by many metrics.
The metric $L_\infty(u,v)=\max\limits_{i=1,\dots,k}|\vb*{u}_i-\vb*{v}_i|$ for any vectors $\vb*{u},\vb*{v}\in\R^k$ preserves the intuition of atomic displacements in the following sense. 
If $S$ is obtained from $Q$ by perturbing every point up to a small $\ep$, 
then $L_\infty(\AMD(S;k),\AMD(Q;k))\leq 2\ep$ by \cite[Theorem~9]{widdowson2022average}.
Other distances such as Euclidean can be considered but will accumulate a larger deviation depending on $k$.
\smallskip

All invariants above and metrics on them are measured in the same units as original coordinates, i.e. in Angstroms for crystals given by Crystallographic Information Files (CIFs). 
The Point Packing Coefficient $\PPC(S)$ was defined as the cube root of the cell volume per atom (of the same radius $1\angstrom$) and can be interpreted as an average radius of balls `packed' in a unit cell.
So $\PPC(S)$ is roughly inversely proportional to the physical density but they are exactly related only when materials have the same average atomic mass (total mass of atoms in a unit cell divided by the cell volume). 
\smallskip

\begin{figure}[h!]
\caption{The averages of $\ADA_k$ and standard deviations (1 sigma shaded) vs $\sqrt[3]{k}$ for four databases. 
}
\includegraphics[width=\textwidth]{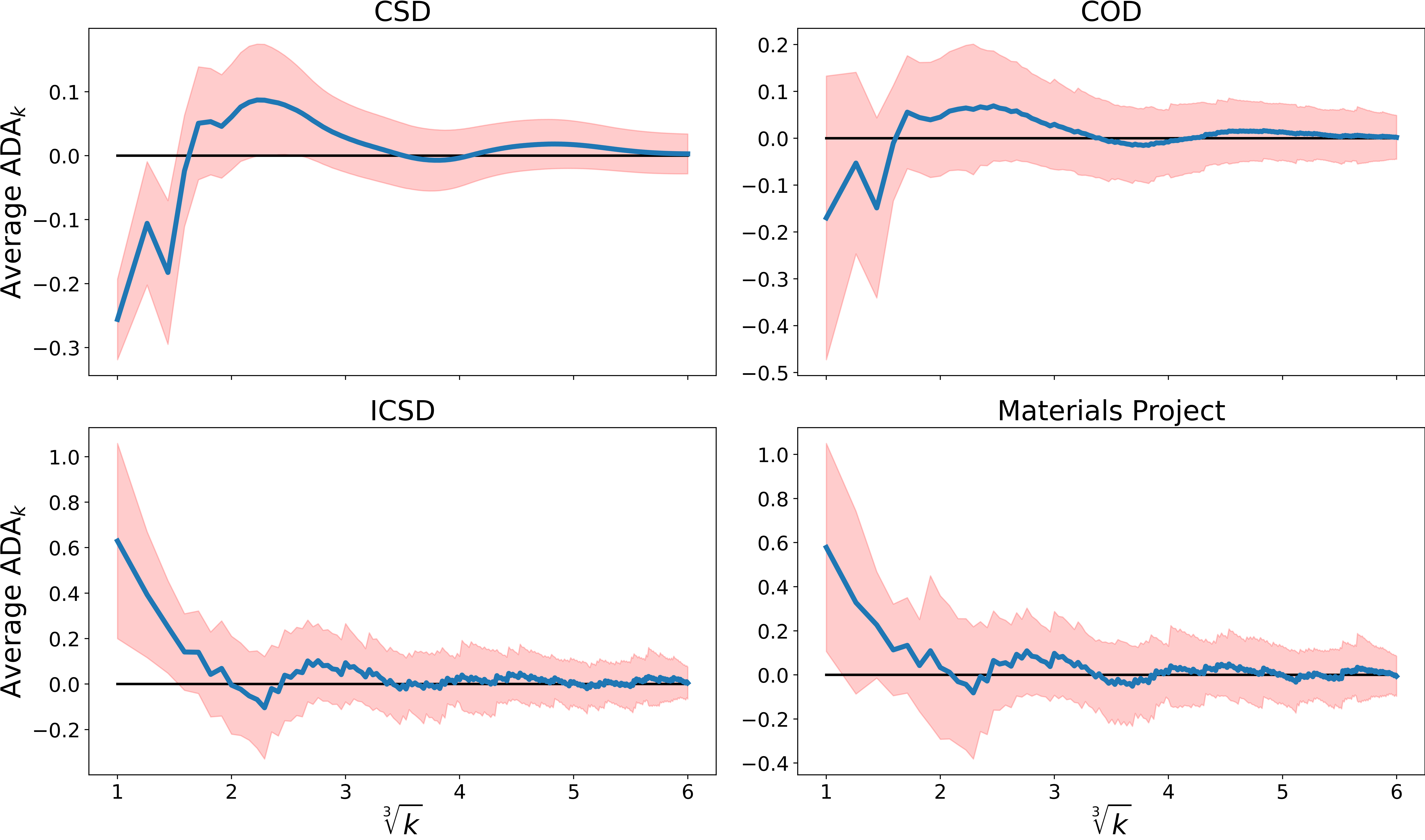}
\label{fig:ADA_averages}
\end{figure}

While $\AMD_k(S)$ monotonically increases in $k$, the invariants $\ADA_k(S)$ can be positive or negative as deviations around the asymptotic $\PPC(S)\sqrt[3]{k}$.
Fig.~\ref{fig:ADA_averages} reveals geometric differences between the mainly organic databases CSD and Crystallography Open Database (COD) \cite{gravzulis2012crystallography} versus the more inorganic collections ICSD and  MP.

The first average of $\ADA_1\in[-0.25,-0.17]$ in the top images of Fig.~\ref{fig:ADA_averages} can be explained by the presence of many hydrogen atoms, which have distances smaller than $\PPC(S)$ to their first neighbor in most organic materials.
Indeed, hydrogens are usually bonded at distances less than $1.2\angstrom$, while $\PPC(S)$ is often larger than $1.2\angstrom$ because most chemical elements have van der Waals radii above $1.2\angstrom$ \cite{batsanov2001van}.
\smallskip

For inorganic materials, metal atoms or ions have relatively large distances to their first neighbors, so the average $\ADA_1$ is in $[0.58,0.62]$ in the bottom images of Fig.~\ref{fig:ADA_averages}.
\smallskip

 
If we increase $k$, the matrix $\PDD(S;k)$ and hence the vector $\ADA(S;k)$ become longer by including distance data to further neighbors but all initial values remain the same.
Hence we consider $k$ not as a parameter that changes the output but as a degree of approximation similarly to the number of decimal places on a calculator.
\smallskip
 
The experimental convergence $\ADA_k\to 0$ as $k\to+\infty$ in Fig.~\ref{fig:ADA_averages} justifies computing the distance $L_\infty$ between $\ADA$ vectors up to a reasonable $k$.
We use $k=100$ because all $\ADA_k$ for $k>100$ are close to 0 (the range of 1 sigma between $\pm 0.2\angstrom$) in Fig.~\ref{fig:ADA_averages}.

\subsection{Novelty distance based on practically complete invariants}
\label{sub:novelty}

This subsection introduces the Local Novelty Distance $\LND(S;D)$ of a periodic crystal $S$ as a distance to the closest neighbor $Q$ of $S$ in a given dataset $D$.
The $\LND$ will be measured as a distance between $\PDA(S;k)$ and $\PDA(Q;k)$ for $Q\in D$.
\smallskip

We can compare PDD matrices that have the same number of columns and possibly different numbers of rows by interpreting $\PDD(S;k)$ as a distribution of unordered rows (or points in $\R^k$) with weights or probabilities.
{
Many similarities between discrete distributions such as the Cramer - von Mises distance and the Kullback - Leibler (KL) divergence fail the axioms of a metric, which are prerequisites for convergence guarantees. 
If the triangle axiom fails with any positive error, outputs of widely used clustering algorithms such as $k$-means and DBSCAN may not be trustworthy} \cite{rass2024metricizing}.
\smallskip

{Though the KL divergence can be symmetrized to the Jensen-Shannon divergence whose square root becomes a metric} \cite{endres2003new}, {these divergences work best for distributions on the same finitely many discrete values, while PDD matrices of different crystals are unlikely to share any common rows (points in $\R^k$) consisting of $k$ continuous distances.} 
\smallskip

{We use the Earth Mover's Distance (EMD) on PDDs, see Definition}~\ref{dfn:EMD} {in the appendix because $\EMD$ is invariant under permutations of rows in $\PDD$s and EMD satisfies all metric axioms} \cite[Appendix]{rubner2000earth}.
{The EMD can be extended to more complicated Wasserstein metrics} \cite{givens1984class} {but the simplest EMD behaves most nicely under bounded noise, motivated by atomic vibrations. 
If any point is perturbed up to $\ep$, any inter-point distance (a value in PDD) can become smaller or larger but only up to $2\ep$, which will allow us to prove the same upper bound $2\ep$ for $\EMD$ in Theorem}~\ref{thm:LND}. 
\smallskip

\begin{dfn}[Local Novelty Distance $\LND(S;D)$]
\label{dfn:LND}
Let $D$ be a finite dataset of periodic point sets.
Fix an integer $k\geq 1$.
For any periodic point set $S$, the \emph{Local Novelty Distance} $\LND(S;D)=\min\limits_{Q\in D} \EMD(\PDA(S;k),\PDA(Q;k))$ is the shortest $L_\infty$-based $\EMD$ distance from $S$ to its nearest neighbor $Q$ in the given {crystal} dataset $D$. 
\end{dfn}
\smallskip

If $S$ is already contained in the dataset $D$, then $\LND(S;D)=0$, so $S$ cannot be considered novel.
{Conversely, if $\LND(S;D)=0$ then $S$ highly likely belongs to $S$, because $\PDD(S;100)$ distinguished all non-duplicate periodic crystals in the CSD.}
\smallskip

Also, for a generic periodic set $S$ (away from a measure 0 subspace), $\PDD(S;k)$ with a big enough $k$ and a lattice of $S$ suffices to reconstruct $S$ uniquely under isometry in $\R^n$ by \cite[Theorem~4.4]{widdowson2022resolving}.
{$\LND(S;D)$ is based on $\PDA$s instead of $\PDD$s because} distances to $k$-th neighbors in $\PDD(S;k)$ asymptotically increase as $\PPC(S)\sqrt[3]{k}$ by \cite[Theorem~4.4]{widdowson2022average}.
If crystals $S,Q$ have $\PPC(S)\neq\PPC(Q)$, the distance $L_\infty$ between rows of $\PDD$s equals the largest {absolute difference of $i$-th distances, which likely happens for $i=k$.} 
So subtracting $\PPC(S)\sqrt[3]{k}$ in Definition~\ref{dfn:PDA} makes any metric on $\PDA$s more informative than on $\PDD$s.
If a newly synthesized periodic crystal $S$ is a near-duplicate of some known {$Q\in D$, then} $\LND(S;D)$ is small as justified below.
The \emph{packing radius} $r(Q)$ is the minimum half-distance between any points of $Q$.
\smallskip
 
\begin{thm}
\label{thm:LND}
If $S$ is obtained from a crystal $Q$ in a dataset $D$ by perturbing every point of $Q$ up to $\ep<r(Q)$, then $\LND(S;D)\leq2\ep$.
To get $S$ from a crystal $Q\in D$ with $\LND(S;D)<2r(Q)$, some atom of $Q$ should be perturbed by at least $0.5\LND(S;D)$.
\end{thm}
\smallskip

Theorem~\ref{thm:LND} is proved in Appendix~A. 
The distance $\LND(S;D)$ is called \emph{local} because Definition~\ref{dfn:LND} uses the first nearest neighbor of $S$ in $D$.
Another novelty of $S$ can be characterized with respect to a global distribution of all crystals in $D$, which we will explore in a forthcoming work.
The local novelty is more urgently needed to tackle the growing crisis of duplication in experimental and simulated databases, some of which were publicly rebutted in  \cite{leeman2024challenges,chawla2024crystallography,juelshol2025continued}, and \cite{cheetham2024artificial}, \cite[Tables~1-2 in section~6]{anosova2024importance}, respectively.

\subsection{Insufficiency of past invariants and similarities of crystals}
\label{sub:similarities}

This subsection briefly reviews the past approaches to classify crystals. 
Some widely used similarities such as the Root Mean Square Deviation (RMSD) \cite{chisholm2005compack} deserve their own detailed discussions in another forthcoming work.
{Conventional settings were thoroughly developed to uniquely represent any periodic crystal in a reduced cell} \cite{parthe2013typix} and can be theoretically considered complete under rigid motion but discontinuously change under almost any perturbation of atoms in practice as shown in Fig.~\ref{fig:perturbations}.
\smallskip

Indeed, perturbations in Fig.~\ref{fig:perturbations} apply to any crystal and can arbitrarily extend a reduced cell to a larger cell whose size cannot be reduced.
Searching for a small perturbation (pseudo-symmetry) to make a cell smaller \cite{zwart2008surprises} inevitably uses thresholds and leads to a trivial classification due to the transitivity axiom, see \cite[section~1]{anosova2024importance}.  
\smallskip

The COMPACK algorithm \cite{chisholm2005compack} outputs an RMSD quantity by comparing finite portions of only molecular crystals.
Its implementation in Mercury also uses thresholds for acceptable deviations of atoms and angles. 
Even if these thresholds are ignored (made large), the algorithm chooses one molecule in a unit cell and 14 (by default) closest molecules around it.
The resulting molecular group depends on a central molecule. 
Even for simple crystals based on a single molecule as is often the case in Crystal Structure Prediction \cite{pulido2017functional}, the choice of 14 (or any other number of) neighbors can be discontinuous when a central molecule has 14th and 15th neighbors at the same distance. 
Selected clusters of molecules in two crystals require an optimal alignment, which is a hard problem because atomic sets can contain numerous indistinguishable atoms, so the optimization must consider many potential permutations.
This problem of exponentially many permutations was resolved in \cite{widdowson2023recognizing} but a choice of a single atomic environment in a periodic crystal remains discontinuous {as shown in} Fig.~\ref{fig:perturbations}.
\smallskip

Other similarities based on all atomic environments such as SOAP \cite{bartok2013representing} and MACE \cite{kovacs2023evaluation} use a Gaussian deviation and a cut-off radius for interatomic interactions to convert a periodic set of discrete points to a complicated smooth function.
This function decomposes into an infinite sum of spherical harmonics whose truncation up to a finite order can become incomplete. 
Appendix~A explains why the structure matcher from pymatgen \cite{pymatgen} can classify many near-duplicates as ``unique crystal structures''.  
\smallskip

The PXRD similarity compares crystals through powder diffraction patterns that are identical for all homometric structures \cite{patterson1939homometric}, some of which were distinguished even by $\AMD_2$ in \cite[appendix~A]{widdowson2022average}.
The PXRD as implemented in Mercury \cite{macrae2020mercury} also fails the triangle inequality but runs faster than the RMSD and SOAP similarities.
\smallskip

In summary, the past approaches through conventional representations and environment-based similarities separately focused on two important complementary properties: completeness and continuity.
The problem of combining these two properties was first stated in \cite{mosca2020voronoi} for lattices and then extended in \cite{anosova2021isometry} to a complete invariant isoset of any periodic point set 
and a continuous metric approximated with a small error factor by an algorithm whose time polynomially depends on the motif size \cite{anosova2026recognition}. 

\section{Results: novelty of materials and navigation maps}
\label{sec:results}

This section describes how the 43 materials reported by A-lab can be automatically  
positioned relative to the ICSD and MP within the full materials space $\CRIS(\R^3)$.
\smallskip

Among the 43 materials whose CIFs are available in the supplementary materials in \cite{szymanski2023autonomous}, only 32 are pure periodic without any disorder, 
10 have \emph{substitutional} disorder with one or more sites occupied by multiple atomic types, and one has \emph{positional} disorder with an atom occupying any of 4 positions with occupancy 0.5.
\smallskip

Closest neighbors within the ICSD and Materials Project for each A-lab crystal were found as follows. 
Using binary search on $\ADA(S;100)$ vectors with the metric $L_\infty$, we found the nearest 100 neighbors for each A-lab crystal within each database. 
These neighbors were then re-compared by Earth Mover's Distance on the stronger invariants $\PDA(S;100)$.
This $\EMD$ metric also outputs which atomic types and/or occupancies were correctly matched and which were not.
Since most A-lab crystals had several nearest neighbors with small distances $\EMD$, we selected the neighbor with the most similar composition as measured by element mover's distance \cite{hargreaves2020earth} in Tables~\ref{tab:A-lab-ICSD-matches} and~\ref{tab:A-lab-MP-matches} below. 
The local novelty distance of each A-lab crystal is not more than the Earth Mover's Distance listed in the column $\EMD_{100}$.
All experiments were run on a desktop computer: AMD Ryzen 5 5600X (6-core), 32GB RAM, Python 3.9, see the Python code with instructions and examples in the supplementary materials.
Table~\ref{tab:neighbor-times} {shows running times, see a linear-time asymptotic of neighbor search in} \cite{elkin2023new}.
\smallskip

\begin{table}[!h]
\begin{tabular}{l|ll}
Stage & ICSD (s) & MP (s) \\ \hline
Binary search on $\ADA(S;100)$ in the full database & 3.023 & 2.450 \\
$\PDA(Q;100)$ for 100 neighbors $Q$ of $S$ found by $\ADA$ & 5.272 & 5.990 \\
$\EMD$ on PDAs for 100 neighbors $Q$ found by $\ADA$ & 0.535 & 0.742 \\
Elemental Mover's Distance (ElMD) for 100 neighbors & 9.534 & 9.737
\end{tabular}
\caption{Time (seconds) to complete each stage of the process of finding nearest neighbors in the ICSD and Materials Project for 43 A-lab crystals \cite{szymanski2023autonomous} on a modest desktop computer.
{The binary search used 6-cores for multiprocessing}.}
\label{tab:neighbor-times}
\end{table}


\begin{table}[!h]
	\begin{tabular}{p{2.65cm}p{1.12cm}p{2.9cm}p{0.9cm}p{3.7cm}}
		A-lab name               & ICSD ID & ICSD composition                                   & $\EMD_{100}$ & Site mismatches                                                                           \\ \hline
		\ce{Ba2ZrSnO6}*          & 181433  & \ce{In_{0.5}Nb_{0.5}BaO3}                          & 0.003       & \ce{Zr_{0.5}Sn_{0.5}} $\leftrightarrow$ \ce{Nb_{0.5}In_{0.5}}                             \\
		\ce{Ba6Na2Ta2V2O17    }  & 97524   & \ce{Ba6Na2Ru2V2O17}                                & 0.092       & Ta $\leftrightarrow$ Ru                                                                   \\
		\ce{Ba6Na2V2Sb2O17    }  & 97524   & \ce{Ba6Na2Ru2V2O17}                                & 0.081       & Sb $\leftrightarrow$ Ru                                                                   \\
		\ce{Ba9Ca3La4(Fe4O15)2}* & 72336   & \ce{Ca3La4Fe8Ba9O30}                               & 0.192       & \ce{(Ca_{0.43}La_{0.57})2Ba} $\leftrightarrow$ \ce{Ca_{0.33}La_{0.67}(Ca_{0.5}Ba_{0.5})2} \\
		\ce{CaCo(PO3)4        }  & 300027   & \ce{Co2P4O12}                                     & 0.172       & Ca $\leftrightarrow$ Co                                                           \\
		\ce{CaFe2P2O9         }  & 79735   & \ce{CaV2P2O9}                                     & 0.073       & Fe $\leftrightarrow$ V                                                            \\
		\ce{CaGd2Zr(GaO3)4    }* & 202850  & \ce{Ca_{0.95}Zr_{0.95}Gd_{2.05}} \ce{Ga_{4.05}O12} & 0.123       & \ce{GaZrGdCa} $\leftrightarrow$ \ce{Ga_{0.52}Zr_{0.48}Ca_{0.32}Gd_{0.68}}                 \\
		\ce{CaMn(PO3)4        }  & 412558  & \ce{MnP2O6}                                        & 0.132       & \ce{Ca} $\leftrightarrow$ \ce{Mn}                                                         \\
		\ce{CaNi(PO3)4        }  & 37136   & \ce{NiCoP4O12}                                     & 0.204       & \ce{Ca} $\leftrightarrow$ \ce{Co}                                                         \\
		\ce{FeSb3Pb4O13       }* & 65839   & \ce{CrSb3Pb_{3.93}O13  }                           & 0.086       & \ce{Fe_{0.25}} $\leftrightarrow$ \ce{Cr_{0.25}}                                           \\
		\ce{Hf2Sb2Pb4O13      }  & 84759   & \ce{W_{4.48}Sn_{11.5}Pb_{15.8}O_{51.9}}            & 0.086       & SbHf $\leftrightarrow$ \ce{Sn_{0.72}W_{0.28}}                                             \\
		\ce{InSb3(PO4)6       }  & 72735  & \ce{Sb2P3O12}                                    & 0.193       & In $\leftrightarrow$ Sb                                         \\
		\ce{InSb3Pb4O13       }  & 49531  & \ce{Pb2Ru2O_{6.5}}                                      & 0.147       & \ce{SbIn} $\leftrightarrow$ \ce{Ru}                                             \\
		\ce{K2TiCr(PO4)3      }  & 40307   & \ce{K2P3Ti2O12}                                   & 0.098       & \ce{Cr} $\leftrightarrow$ \ce{Ti}                     \\
		\ce{K4MgFe3(PO4)5     }  & 263040   & \ce{Fe4K4P5O20}                                  & 0.139       & Mg $\leftrightarrow$ \ce{Fe}                                            \\
		\ce{K4TiSn3(PO5)4     }  & 79650   & \ce{KPSnO5} & 0.094      & \ce{Ti} $\leftrightarrow$  \ce{Sn}             \\
		\ce{KBaGdWO6          }  & 60499   & \ce{WCaBa2O6}                                      & 0.009       & GdK $\leftrightarrow$ \ce{CaBa}                                                           \\
		\ce{KBaPrWO6          }  & 60499   & \ce{WCaBa2O6}                                      & 0.053       & PrK $\leftrightarrow$ \ce{CaBa}                                                           \\
		\ce{KMn3O6            }* & 261406  & \ce{K_{0.463}MnO2}                                 & 0.016       & \ce{K_{0.5}} $\leftrightarrow$ \ce{K_{0.695}}                                             \\
		\ce{KNa2Ga3(SiO4)3    }  & 411328  & \ce{SiNaGaO4}                                      & 0.27       & SiGaK $\leftrightarrow$ GaSiNa                                                            \\
		\ce{KNaP6(PbO3)8      }* & 182501  & \ce{KNaP6Pb8O24}                                   & 0.005       &                                                                                           \\
		\ce{KNaTi2(PO5)2      }  & 68705    & \ce{KPTiO5}                        & 0.157       & Na $\leftrightarrow$ K                                              \\
		\ce{KPr9(Si3O13)2     }* & 153272  & \ce{KSi6Pr9O26}                                    & 0.16       & \ce{(K_{0.1}Pr_{0.9})2} $\leftrightarrow$ \ce{PrK_{0.25}Pr_{0.75}}                        \\
		\ce{Mg3MnNi3O8        }  & 40584    & \ce{MnNi6O8}                                    & 0.020       & Mg $\leftrightarrow$ Ni                                             \\
		\ce{Mg3NiO4           }* & 109086    & \ce{TaLi3O4}                             & 0.000       & \ce{Mg_{0.75}Ni_{0.25}} $\leftrightarrow$ \ce{Li_{0.75}Ta_{0.25}}                           \\
		\ce{MgCuP2O7          }* & 69576   & \ce{Co_{0.92}Mg_{1.08}P2O7}                        & 0.218       & \ce{Mg_{0.5}Cu_{0.5}} $\leftrightarrow$ \ce{Mg_{0.54}Co_{0.46}}                           \\
		\ce{MgNi(PO3)4        }  & 37137    & \ce{NiZnP4O12}                                     & 0.132       & Mg $\leftrightarrow$ Zn                                                                   \\
		\ce{MgTi2NiO6         }  & 171584   & \ce{NiTiO3}                                     & 0.047  & Mg $\leftrightarrow$ \ce{Ni} \\
		\ce{MgTi4(PO4)6       }  & 419418  & \ce{MnTi4P6O24}                                    & 0.133  & Mg $\leftrightarrow$ Mn                                                                   \\
		\ce{MgV4Cu3O14        }  & 164189    & \ce{Cu2V2O7}                                    & 0.146  & Mg $\leftrightarrow$ \ce{Cu}                                            \\
		\ce{Mn2VPO7           }  & 20296   & \ce{Mn2P2O7         }                              & 0.21  & V $\leftrightarrow$ P                                                                     \\
		\ce{Mn4Zn3(NiO6)2     }  & 625     & \ce{MgCu2Mn3O8}                                    & 0.186  & MnZnNi $\leftrightarrow$ MgCuMn                                                           \\
		\ce{Mn7(P2O7)4        }  & 67514   & \ce{Fe7P8O28        }                              & 0.126  & Mn $\leftrightarrow$ Fe                                                                   \\
		\ce{MnAgO2            }  & 670065  & \ce{MnAgO2          }                              & 0.097  &                                                                                           \\
		\ce{Na3Ca18Fe(PO4)14  }  & 85103   & \ce{FeNa3P14Ca18O56}                               & 0.153  & \ce{FeCa2Na} $\leftrightarrow$ \ce{Ca_{0.5}Fe_{0.5}Na_{0.17}Ca_{0.83}}                    \\
		\ce{Na7Mg7Fe5(PO4)12  }  & 200238  & \ce{Na2Fe3P3O12     }                              & 0.229  & \ce{POMg2} $\leftrightarrow$ \ce{Na3Fe}                                                   \\
		\ce{NaCaMgFe(SiO3)4   }* & 172120  & \ce{NaCaMgCrSi4O12  }                              & 0.075  & \ce{(MgFeNaCa)_{0.25}} $\leftrightarrow$ MgCrNaCa                                         \\
		\ce{NaMnFe(PO4)2      }  & 200238  & \ce{Na2Fe3P3O12     }                              & 0.242  & \ce{POMn2} $\leftrightarrow$ \ce{Na2Fe2}                                                  \\
		\ce{Sn2Sb2Pb4O13      }  & 49533   & \ce{PbNb2Tl_{0.9}O_{6.45}}                         & 0.209  & \ce{SbSnPb} $\leftrightarrow$ \ce{NbTl}                                               \\
		\ce{Y3In2Ga3O12       }  & 185862  & \ce{Y3Ga5O12}                                      & 0.104  & In $\leftrightarrow$ Ga                                                                   \\
		\ce{Zn2Cr3FeO8        }  & 196119  & \ce{ZnCr2O4}                                       & 0.022  & Fe $\leftrightarrow$ Cr                                                                   \\
		\ce{Zn3Ni4(SbO6)2     }* & 180711  & \ce{Ti_{0.18}Zr_{0.33}ZnO2}                        & 0.162  & \ce{Ni_{0.66}Sb_{0.33}} $\leftrightarrow$ \ce{Ti_{0.17}Zn_{0.5}Zr_{0.33}}                 \\
		\ce{Zr2Sb2Pb4O13      }  & 65054   & \ce{TiSbPb_{1.97}O_{6.5}}                          & 0.12  & SbZr $\leftrightarrow$ \ce{Ti_{0.5}Sb_{0.5}}
	\end{tabular}
	\caption{Close neighbors of each A-lab crystal in the ICSD. 
		The ICSD entry with the smallest element mover's distance \cite{hargreaves2020earth} was selected from the list of 100 nearest neighbors by $\ADA_{100}$. Disordered crystals are marked with an asterisk *.
	}
     \label{tab:A-lab-ICSD-matches}
\end{table}

The GNoME (Graph Network Materials Exploration) \cite{merchant2023scaling} was trained on a snapshot of the Materials Project database (whose entries are partly sourced from the ICSD) from 2021 and made public 384,938 crystals. 
Berkeley's A-lab attempted to synthesize 58 of them and reported 43  \cite{szymanski2023autonomous}, which were split into 36 ``successes'' and 7 ``partial successes'' (less than 50\% of the weight of solute versus the weight of solution).
Table~I in \cite{leeman2024challenges} summarized four types of issues for the 36 ``successes'', where only 3 were marked as already reported structures.
Table \ref{tab:A-lab-ICSD-matches} lists geometric close matches that were automatically found in the ICSD for all 43 A-lab crystals within a few seconds.
\smallskip

\begin{figure}[h!]
\caption{\textbf{Left}: \ce{MnAgO2} synthesized by A-lab.
\textbf{Middle}: ICSD entry 670065 with the same composition and $\EMD=0.097\angstrom$ found by structural invariants in Table~\ref{tab:A-lab-ICSD-matches}, though its unit cell is very different from the cell of \ce{MnAgO2}.
\textbf{Right}: another ICSD entry 139006 from 2021 matched by \cite{leeman2024challenges} and found by unit cell search, but is more distant from \ce{MnAgO2} by $\EMD=0.368\angstrom$ on invariants $\PDA(S;100)$.}
\includegraphics[height=40mm]{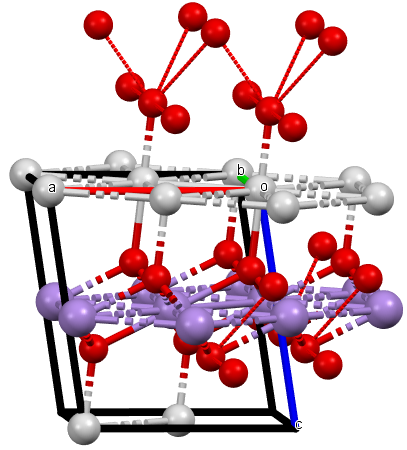}
\includegraphics[height=40mm]{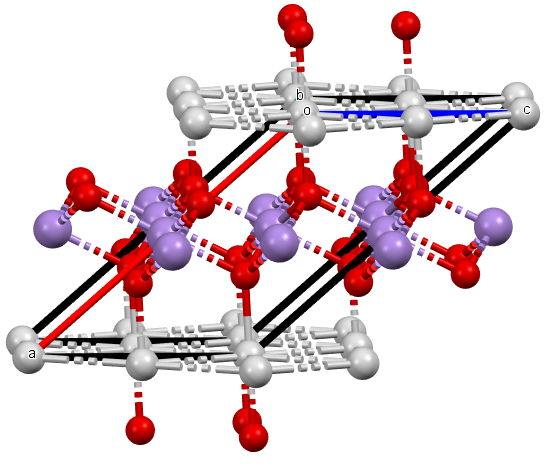}
\includegraphics[height=40mm]{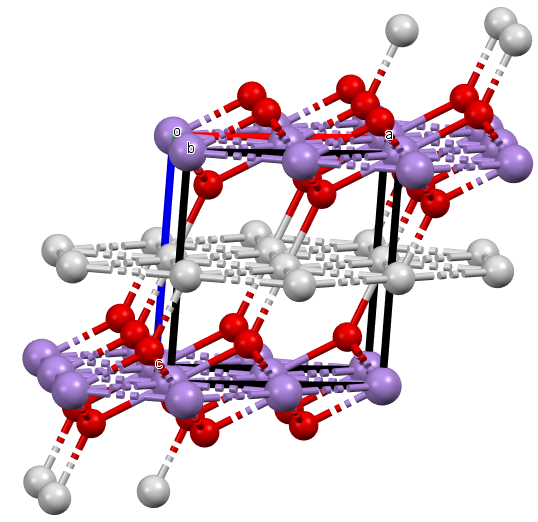}
\label{fig:MnAgO2}
\end{figure}

Two A-lab crystals were found to already exist in the ICSD with the same composition: \ce{KNaP6(PbO3)8} matched ICSD 182501 reported in 2011 \cite{azrour2011rietveld}, and \ce{MnAgO2} matched ICSD 670065 reported as a hypothetical structure in 2015 \cite{doi:10.1021/acs.chemmater.5b00716}.
In particular, \ce{MnAgO2} was one of three crystals that the later rebuttal said was synthesized successfully \cite{leeman2024challenges}, and they go on to state that the material was first reported in 2021 \cite{griesemer2021high} (ICSD 139006), after the snapshot used to train the GNoME, and so was not included in the original training data and could be considered a success. 
Our findings show this crystal did in fact exist in the ICSD prior to the 2021 snapshot.
The pre-existing version of this crystal was not found by \cite{leeman2024challenges} using a unit cell search because the unit cell of ICSD 670065 significantly differs from that of the A-lab version or ICSD 139006, with the former listing its space group as A 2/m and the latter two having space group C 2/m, see Fig.~\ref{fig:MnAgO2}. 
Such cell-based search can always miss near-duplicates as in Fig.~\ref{fig:perturbations}, while continuous invariants independent of a unit cell find near-duplicates despite disagreement on a space group, which breaks down under almost any noise.

\vspace*{-2mm}
\vspace*{-2mm}

Aside from the two structures above, all other A-lab crystals were found to have a geometric near-duplicate in the ICSD with a different composition.
Many of these near-duplicates involve the substitution of only one atom, replacing a disordered site with a fully ordered one or adjusting the occupancy ratios of atoms at a site. 
\smallskip

These structural analogues of A-lab’s reported materials are not surprising as the GNoME AI \cite{merchant2023scaling} used atomic substitution on existing crystals to generate potential new ones without substantially changing the atomic geometry.
The fact that pre-existing structures in the ICSD were missed by the later rebuttal \cite{leeman2024challenges} suggests that a more robust method is needed for comparing structures in the aid of materials discovery.
\smallskip


The Materials Project contains {numerous theoretical structures, many of which are obtained by substituting atoms in experimental} structures with plausible alternatives, a strategy also employed by the GNoME which generated the crystals later synthesized by Berkeley's A-lab. 
Despite the substitution patterns used by GNoME being tuned to prioritize discovery and not repeat data, 42 of the 43 A-lab crystals were found to already exist in the Materials Project, all of which predate the March 2021 snapshot used to train the GNoME and hence were part of its training data. 
\smallskip

As the Materials Project does not model disorder, no match was found for the positionally disordered \ce{KMn3O6}. 
However, its nearest neighbor was found in the ICSD with a change in occupancy.
So all 43 A-lab crystals had already been hypothesized or synthesized prior to the beginning of the GNoME project, see Table~\ref{tab:experimental-matches}. 
\smallskip

\begin{table}[!h]
\begin{tabular}{lll}
A-lab name & Matching database entries & Source and date \\ \hline
\ce{Ba6Na2Ta2V2O_{17}} & mp-1214664, Pauling file sd\_1003187 & \cite{quarez2003synthesis}, 2003 \\
\ce{Ba6Na2V2Sb2O_{17}} & mp-1214658, Pauling file sd\_1003189 & \cite{quarez2003synthesis}, 2003 \\
\ce{CaGd2Zr(GaO3)4} & mp-686296, ICSD 202850 & \cite{julien1988structure}, 1988 \\
\ce{KNa2Ga3(SiO4)3} & mp-1211711, Pauling file sd\_1707156 & \cite{selker1982struktur}, 1982 \\
\ce{KNaP6(PbO3)8} & ICSD 182501 & \cite{azrour2011rietveld}, 2011 \\
\ce{KNaTi2(PO5)2} & mp-1211611, Pauling file sd\_1414297 & \cite{crennell1991isomorphous}, 1991 \\
\ce{Mn2VPO7} & mp-1210613, Pauling file sd\_1322766 & \cite{yakubovich2000crystal}, 2000 \\
\ce{Y3In2Ga3O_{12}} & mp-1207946, Pauling file sd\_1704376 & \cite{schmitz1964farbe}, 1964
\end{tabular}
\caption{The eight reportedly new crystals synthesized by the A-lab found to already have been synthesized and uploaded to various databases.}
\label{tab:experimental-matches}
\end{table}

 \begin{table}[!h]
	\begin{tabular}{llllp{3.4cm}}
		A-lab name               & MP ID   & MP composition          & $\EMD_{100}$ & Site mismatches                                              \\ \hline
		\ce{Ba2ZrSnO6         }* & 1228067 & \ce{Ba2ZrSnO6         } & 0.025       & \ce{Zr_{0.5}Sn_{0.5}} $\leftrightarrow$ ZrSn                 \\
		\ce{Ba6Na2Ta2V2O17    }  & 1214664 & \ce{Ba6Na2Ta2V2O17 }    & 0.029       &                                                              \\
		\ce{Ba6Na2V2Sb2O17    }  & 1214658 & \ce{Ba6Na2V2Sb2O17 }    & 0.021       &                                                              \\
		\ce{Ba9Ca3La4(Fe4O15)2}* & 1228537 & \ce{Ba9Ca3La4Fe8O30}    & 0.136       & \ce{Ca_{0.43}La_{0.57}} $\leftrightarrow$ \ce{Ca3La4}        \\
		\ce{CaCo(PO3)4        }  & 1045787 & \ce{CaCoP4O12      }    & 0.090       &                                                              \\
		\ce{CaFe2P2O9         }  & 1040941 & \ce{CaFe2P2O9      }    & 0.114       &                                                              \\
		\ce{CaGd2Zr(GaO3)4    }* & 686296  & \ce{CaGd2ZrGa4O12  }    & 0.069       & Ga $\leftrightarrow$ Zr                                      \\
		\ce{CaMn(PO3)4        }  & 1045779 & \ce{CaMnP4O12      }    & 0.163       &                                                              \\
		\ce{CaNi(PO3)4        }  & 1045813 & \ce{CaNiP4O12      }    & 0.151       &                                                              \\
		\ce{FeSb3Pb4O13       }* & 1224890 & \ce{FeSb3Pb4O13    }    & 0.027       & \ce{Fe_{0.25}Sb_{0.75}} $\leftrightarrow$ \ce{FeSb3}         \\
		\ce{Hf2Sb2Pb4O13      }  & 1224490 & \ce{Hf2Sb2Pb4O13   }    & 0.012       &                                                              \\
		\ce{InSb3(PO4)6       }  & 1224667 & \ce{InSb3P6O24     }    & 0.011       &                                                              \\
		\ce{InSb3Pb4O13       }  & 1223746 & \ce{InSb3Pb4O13    }    & 0.029       &                                                              \\
		\ce{K2TiCr(PO4)3      }  & 1224541 & \ce{K2TiCrP3O12    }    & 0.009       &                                                              \\
		\ce{K4MgFe3(PO4)5     }  & 532755  & \ce{K4MgFe3P5O20   }    & 0.076       &                                                              \\
		\ce{K4TiSn3(PO5)4     }  & 1224290 & \ce{K4TiSn3P4O20   }    & 0.014       &                                                              \\
		\ce{KBaGdWO6          }  & 1523079 & \ce{KBaGdWO6       }    & 0.006       &                                                              \\
		\ce{KBaPrWO6          }  & 1523149 & \ce{KBaPrWO6       }    & 0.012       &                                                              \\
		\ce{KMn3O6            }* & 1223545 & \ce{KMn2O4         }    & 0.439       & Not a match                                                  \\
		\ce{KNa2Ga3(SiO4)3    }  & 1211711 & \ce{KNa2Ga3Si3O12  }    & 0.022       &                                                              \\
		\ce{KNaP6(PbO3)8      }* & 1223429 & \ce{KNaP6Pb8O24       } & 0.174       & \ce{Na_{0.25}K_{0.25}Pb_{0.5}} $\leftrightarrow$ \ce{NaKPb2} \\
		\ce{KNaTi2(PO5)2      }  & 1211611 & \ce{KNaTi2P2O10    }    & 0.012       &                                                              \\
		\ce{KPr9(Si3O13)2     }* & 1223421 & \ce{KPr9Si6O26     }    & 0.009       & \ce{K_{0.1}Pr_{0.9}} $\leftrightarrow$ \ce{KPr9}             \\
		\ce{Mg3MnNi3O8        }  & 1222170 & \ce{Mg3MnNi3O8     }    & 0.029       &                                                              \\
		\ce{Mg3NiO4           }* & 1099253 & \ce{Mg3NiO4        }    & 0.002       & \ce{Mg_{0.75}Ni_{0.25}} $\leftrightarrow$ \ce{Mg3Ni}         \\
		\ce{MgCuP2O7          }* & 1041741 & \ce{MgCuP2O7       }    & 0.093       & \ce{Mg_{0.5}Cu_{0.5}} $\leftrightarrow$ MgCu                 \\
		\ce{MgNi(PO3)4        }  & 1045786 & \ce{MgNiP4O12      }    & 0.018       &                                                              \\
		\ce{MgTi2NiO6         }  & 1221952 & \ce{MgTi2NiO6      }    & 0.009       &                                                              \\
		\ce{MgTi4(PO4)6       }  & 1222070 & \ce{MgTi4P6O24     }    & 0.075       &                                                              \\
		\ce{MgV4Cu3O14        }  & 1222158 & \ce{MgV4Cu3O14     }    & 0.060       &                                                              \\
		\ce{Mn2VPO7           }  & 1210613 & \ce{Mn2VPO7        }    & 0.125       &                                                              \\
		\ce{Mn4Zn3(NiO6)2     }  & 1222033 & \ce{Mn4Zn3Ni2O12   }    & 0.054       &                                                              \\
		\ce{Mn7(P2O7)4        }  & 778008  & \ce{Mn7P8O28       }    & 0.123       &                                                              \\
		\ce{MnAgO2            }  & 996995  & \ce{MnAgO2         }    & 0.098       &                                                              \\
		\ce{Na3Ca18Fe(PO4)14  }  & 725491  & \ce{Na3Ca18FeP14O56}    & 0.031       &                                                              \\
		\ce{Na7Mg7Fe5(PO4)12  }  & 1173791 & \ce{Na7Mg7Fe5P12O48}    & 0.028       &                                                              \\
		\ce{NaCaMgFe(SiO3)4   }* & 1221075 & \ce{NaCaMgFeSi4O12 }    & 0.026       & \ce{(MgFeNaCa)_{0.25}} $\leftrightarrow$ MgFeNaCa            \\
		\ce{NaMnFe(PO4)2      }  & 1173592 & \ce{NaMnFeP2O8     }    & 0.032       &                                                              \\
		\ce{Sn2Sb2Pb4O13      }  & 1219056 & \ce{Sn2Sb2Pb4O13   }    & 0.025       &                                                              \\
		\ce{Y3In2Ga3O12       }  & 1207946 & \ce{Y3In2Ga3O12    }    & 0.008       &                                                              \\
		\ce{Zn2Cr3FeO8        }  & 1215741 & \ce{Zn2Cr3FeO8     }    & 0.014       &                                                              \\
		\ce{Zn3Ni4(SbO6)2     }* & 1216023 & \ce{Zn3Ni4Sb2O12   }    & 0.092       & \ce{Ni_{0.67}Sb_{0.33}} $\leftrightarrow$ \ce{Ni2Sb}         \\
		\ce{Zr2Sb2Pb4O13      }  & 1215826 & \ce{Zr2Sb2Pb4O13   }    & 0.025       &
	\end{tabular}
	\caption{Close neighbors of each A-lab crystal in the Materials Project (MP). 
		In each case, the MP entry with the smallest element mover's distance \cite{hargreaves2020earth} was selected from the list of 100 nearest neighbors by $\ADA_{100}$. Disordered crystals are marked with an asterisk *.
	}
	\label{tab:A-lab-MP-matches}
\end{table}

The rebuttal paper \cite{leeman2024challenges} said that the crystal \ce{Y3In2Ga3O_{12}} in Table~\ref{tab:experimental-matches} {was one of the three new crystals to have been synthesized} 
and provided the reference for this crystal to 2022 \cite{li2022highly}, again leading to the conclusion that the crystal was novel from the perspective of the GNoME AI trained on data from 2021. 
We found that the crystal \ce{Y3In2Ga3O_{12}} was reported in 1964 and uploaded to the Materials Project no later than 2018, and so would have been part of GNoME’s training data.
\smallskip

The 10 substitutionally disordered A-lab crystals had matches in the Materials Project where disordered sites were replaced with multiple fully ordered sites of atoms in the same ratio; e.g. \ce{FeSb3Pb4O13} matching mp-1224890 had a site \ce{Fe_{0.25}Sb_{0.75}} with multiplicity 4 replaced with \ce{FeSb3}. 
For completeness, this is noted in the site mismatches column of Table~\ref{tab:A-lab-MP-matches}, listing all nearest neighbors in the Materials Project.
\smallskip

One pair of note is \ce{CaGd2Zr(GaO3)4} \& mp-686296, which have one atom swapped (Ga $\leftrightarrow$ Zr). 
This Materials Project entry originates from ICSD 202850, listed in Table~\ref{tab:A-lab-ICSD-matches} as the closest neighbor in the ICSD.
The ICSD entry has disorder on the sites where atoms were swapped, whereas the A-lab and Materials Project versions have no disorder. 
We conclude that this crystal is not new, as these atoms could have been swapped to match the A-lab crystal with a different ordering of the disordered ICSD entry.
\smallskip

\begin{figure}[h!]
\caption{
A-lab crystals are in cyan over the heat map of ICSD and MP in invariant
coordinates. 
}
\includegraphics[width=0.49\textwidth]{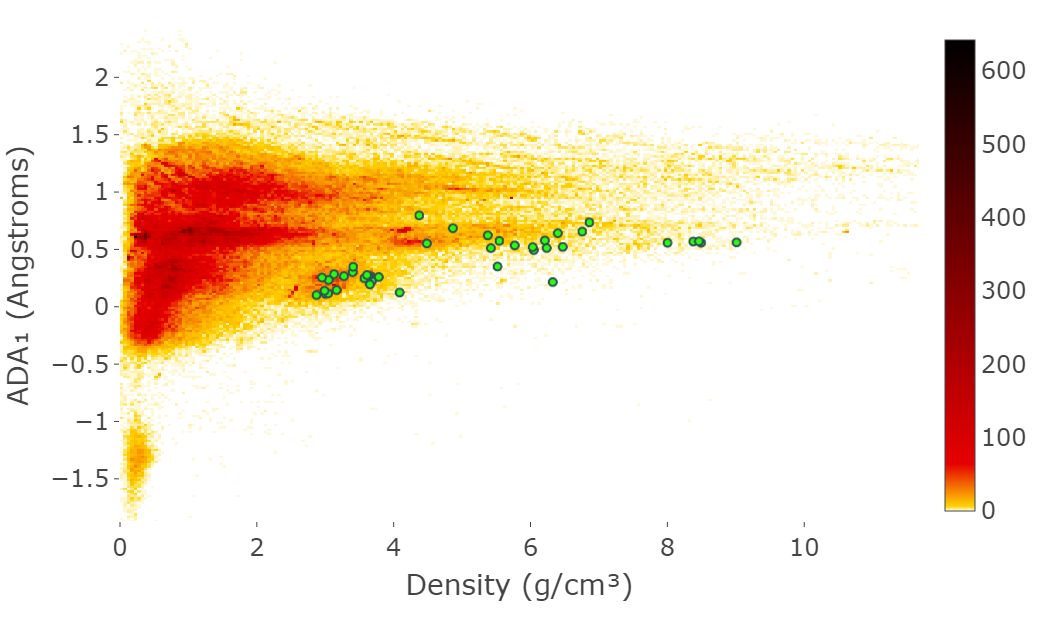}
\includegraphics[width=0.49\textwidth]{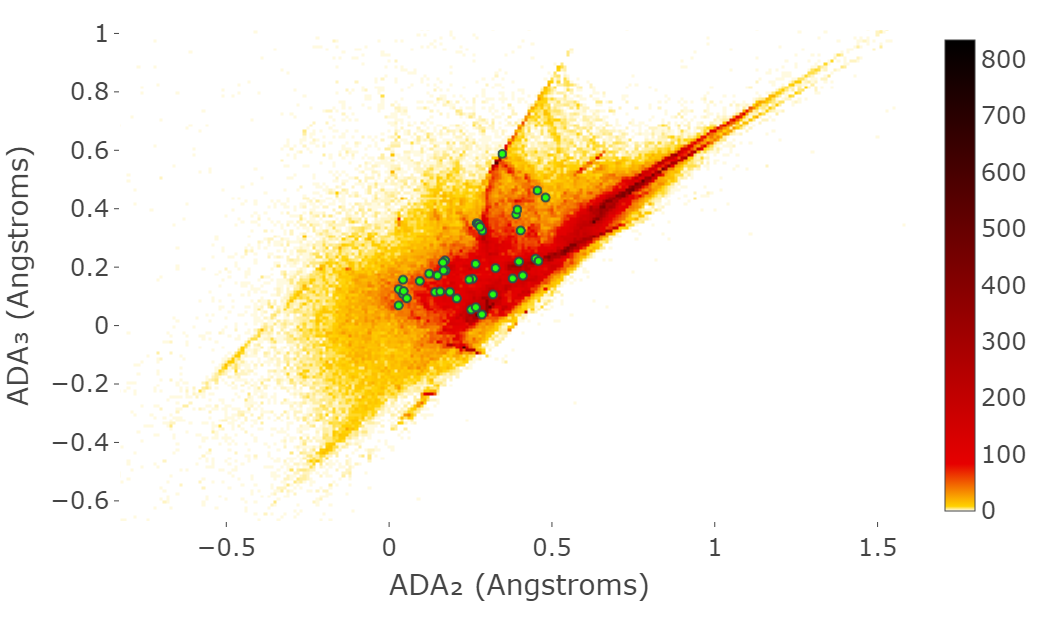}
\includegraphics[width=0.49\textwidth]{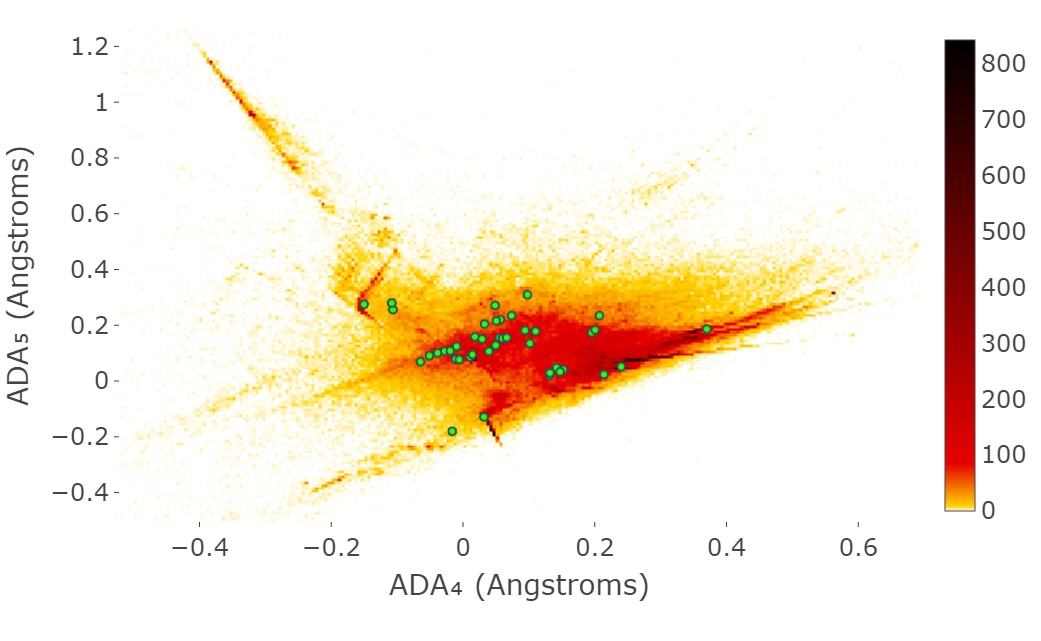}
\includegraphics[width=0.49\textwidth]{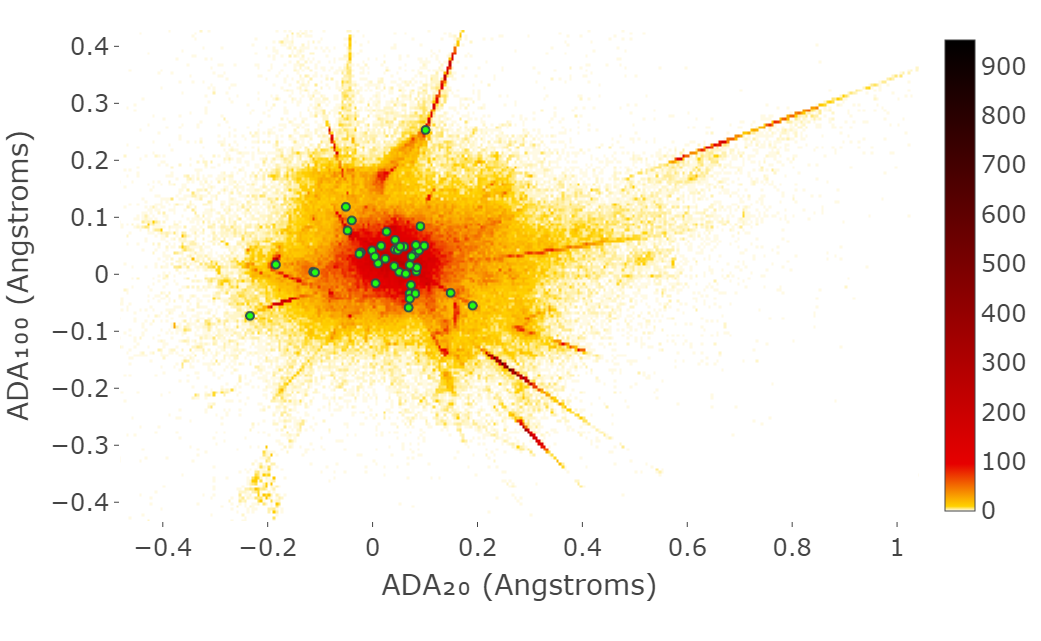}
\label{fig:A-lab_on_ICSD+MP}
\end{figure}

Fig.~\ref{fig:A-lab_on_ICSD+MP} {shows 2D projections (heat maps) of the ICSD and MP to pairs of analytically defined (data-independent) invariant coordinates, see continuous maps of the CSD and its subsets in} \cite{widdowson2024continuous}.
The color of any pixel with coordinates $(x,y)$ indicates the number of crystals whose continuous invariants coincide with $(x,y)$ after discretization to pixels. 
To better visualize hot spots, we excluded some outliers, e.g. all crystals with densities higher than 21 g/cm$^3$. 
{Subspaces of highly symmetric (cubic or primitive orthorhombic) crystals are visible as straight lines due to linear dependencies of inter-atomic distances in these subspaces.
The projections in Fig.}~\ref{fig:A-lab_on_ICSD+MP} {can be considered universal maps of the continuous space of all crystal structures because any newly discovered crystal will appear at its unique location without affecting all known crystals.}
\smallskip

\vspace*{-4mm}

\vspace*{-6mm}
\section{Conclusions: fast navigation in the materials space}
\label{sec:conclusion}

Definition~\ref{dfn:CRIS} formalized the materials universe as the \emph{Crystal Isometry Space} $\CRIS$ containing all known and not yet discovered crystals at unique locations determined by sufficiently precise geometry of atomic centers.
Definition~\ref{dfn:LND} introduced the Local Novelty Distance (LND) based on generically complete invariants of periodic point sets.
The ultra fast $\LND$ quantifies the novelty of any synthesized material as a continuous distance to the nearest crystal structure (independent of chemical composition) from the world's largest databases within seconds on a modest desktop computer.
\smallskip

Tables~\ref{tab:A-lab-ICSD-matches} and~\ref{tab:A-lab-MP-matches} showed that structural near-duplicates of all A-lab crystals existed before the GNoME project and were seemingly part of its training data but were targeted for synthesis.
The recent work \cite{juelshol2025continued} exposed similar challenges of MatterGen \cite{zeni2025generative} using a manual search by chemical composition.
As with car driving, navigation maps based on structural invariants {in Fig.}~\ref{fig:A-lab_on_ICSD+MP} are needed not to get lost in the materials space by quickly recognizing near-duplicates of known crystals without manual work.
\smallskip

The next step in exploring the materials space is to understand the structure-property relations by visualizing property values like mountainous landscapes in Fig.~\ref{fig:CRISP}~(right).  
{The invariant coordinates PDD (generically complete under isometry) helped predict properties of organic and inorganic materials} \cite{ropers2022fast,balasingham2024material,balasingham2024accelerating} {including synthesizability} \cite{schwalbe2023inorganic}.
{Similar maps of the protein universe used linear-time} invariants \cite{anosova2025complete}, which detected thousands of unexpected duplicates in the Protein Data Bank \cite{wlodawer2025duplicate}.
\smallskip

Authors' contributions.
VK developed Definitions~\ref{dfn:CRIS} and~\ref{dfn:LND} and wrote the paper.
DW {implemented the code and} produced all tables.
Both authors reviewed the manuscript.
\smallskip

Acknowledgements.
This work was supported by the EPSRC grant ``Inverse design of periodic crystals'' (EP/X018474/1) and the Royal Society APEX fellowship ``New geometric methods for mapping the space of periodic crystals'' (APX/R1/231152).
We thank Andy Cooper, Robert Palgrave and anonymous reviewers for helpful advice. 
\smallskip

Data Availability.
Data is provided within the 
supplementary information files.

\bibliography{SR_geomaps_arxiv2025Dec}


\newpage

\begin{appendices}

\section{Extra examples and navigation maps}
\label{app:proofs}

\begin{figure}[h!]
\caption{{The invariants $\AMD_k$ and $\ADA_k$ from Definition}~\ref{dfn:PDA} {and their weighted versions taking into account atomic masses for five perovskites from MP in Table}~\ref{tab:MPcrystals}.
\textbf{Left}: unweighted.
\textbf{Right}: weighted.}
\includegraphics[width=0.49\textwidth]{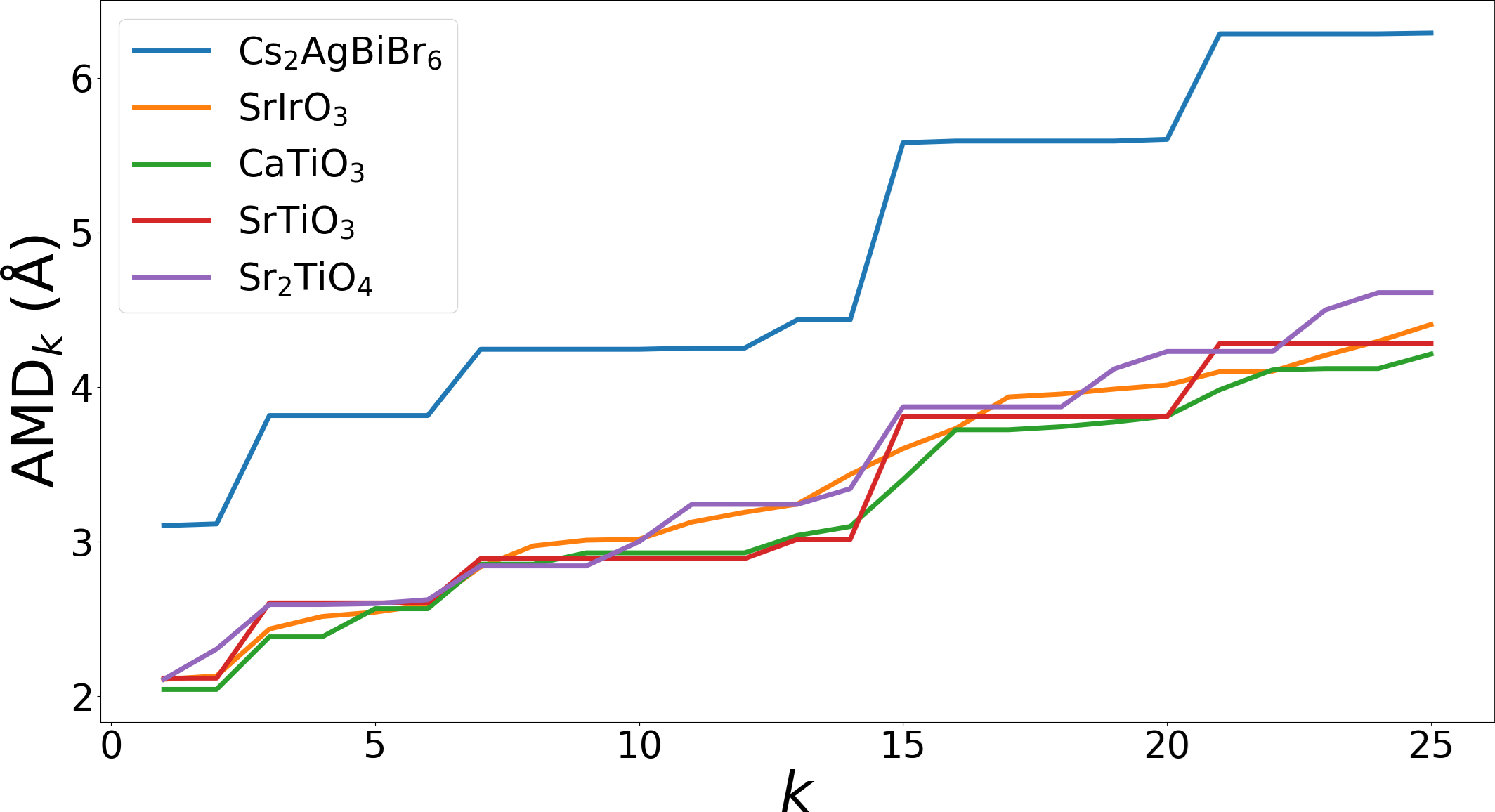}
\includegraphics[width=0.49\textwidth]{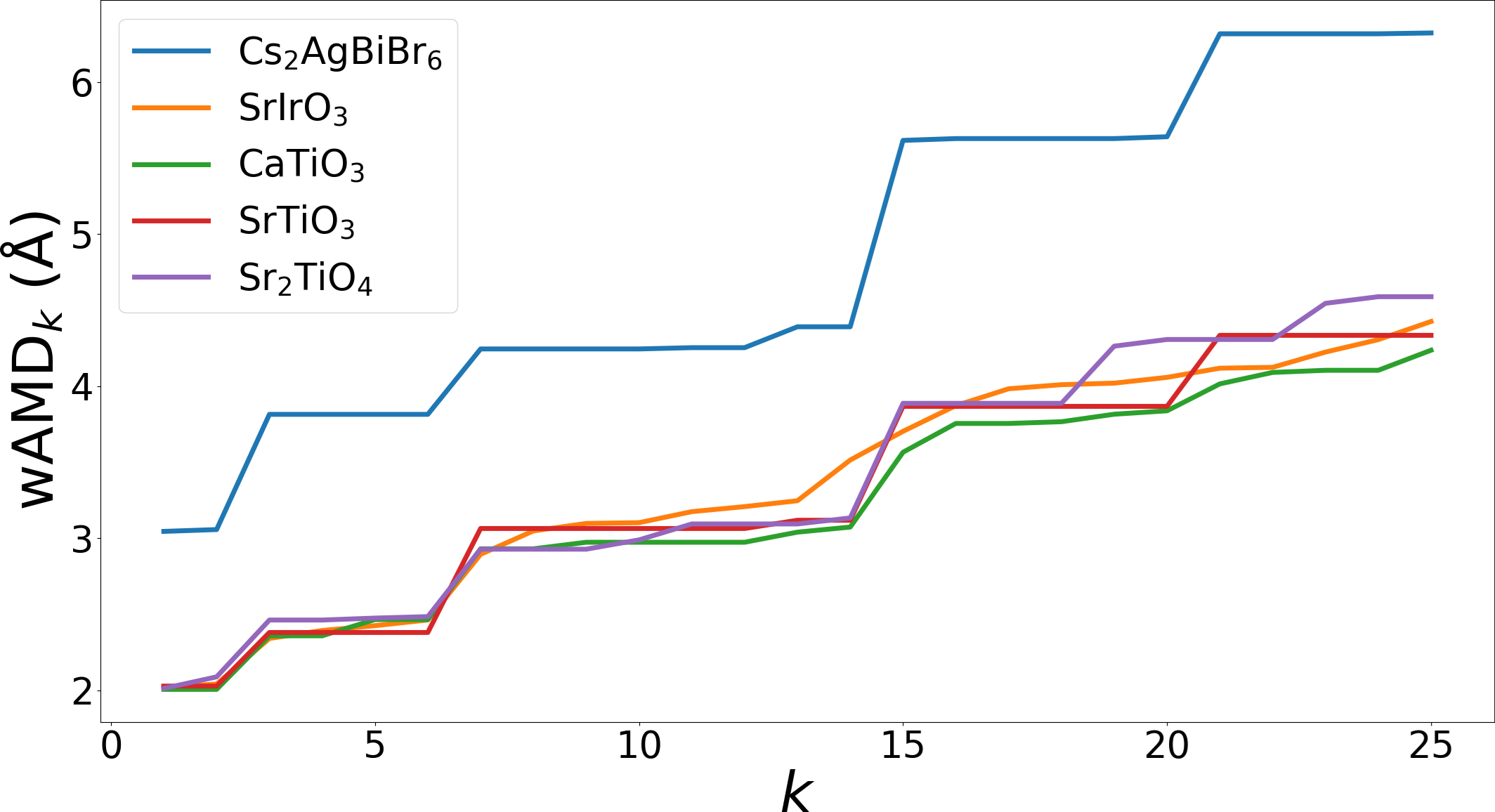}
\smallskip

\includegraphics[width=0.49\textwidth]{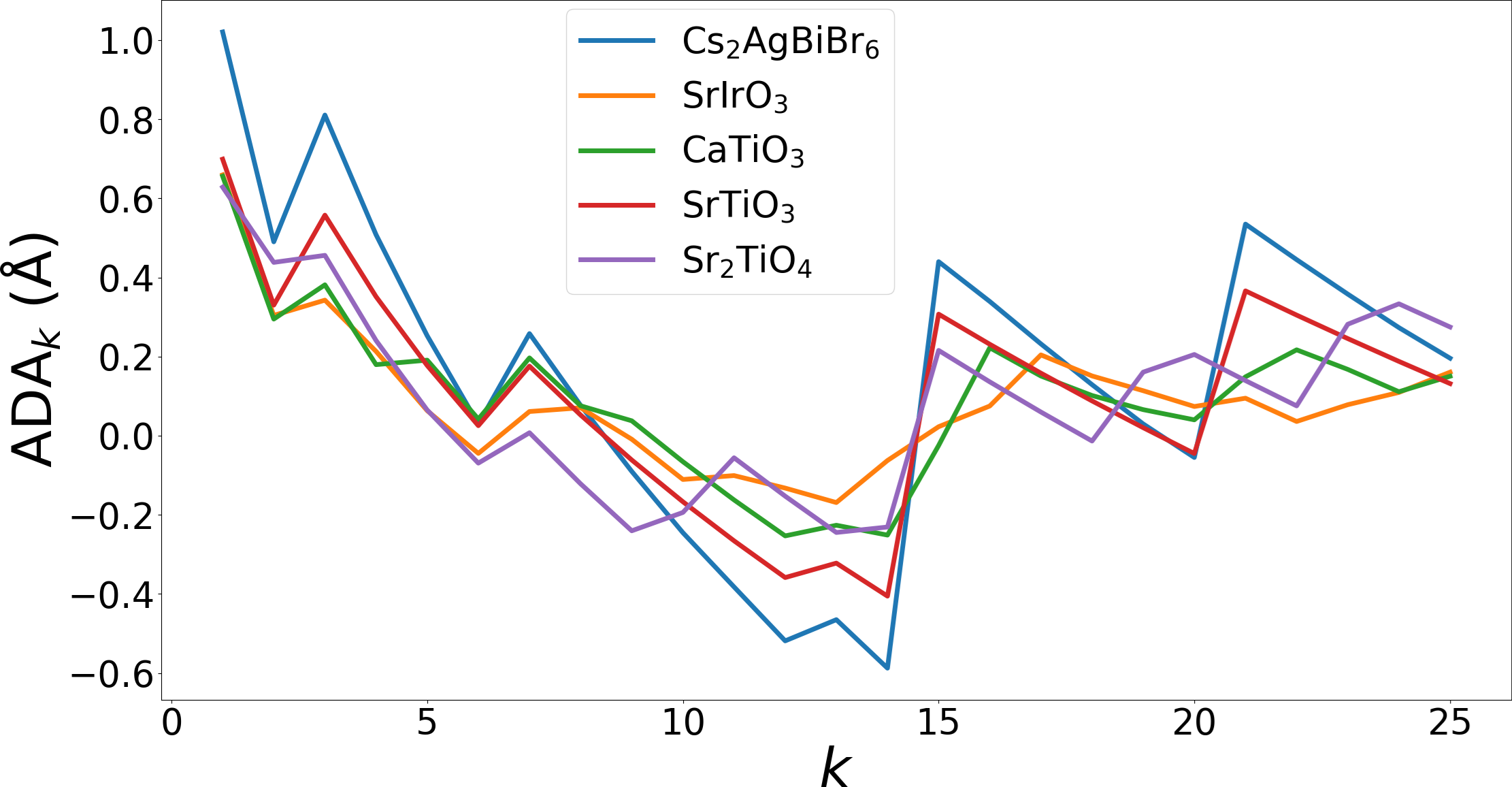}
\includegraphics[width=0.49\textwidth]{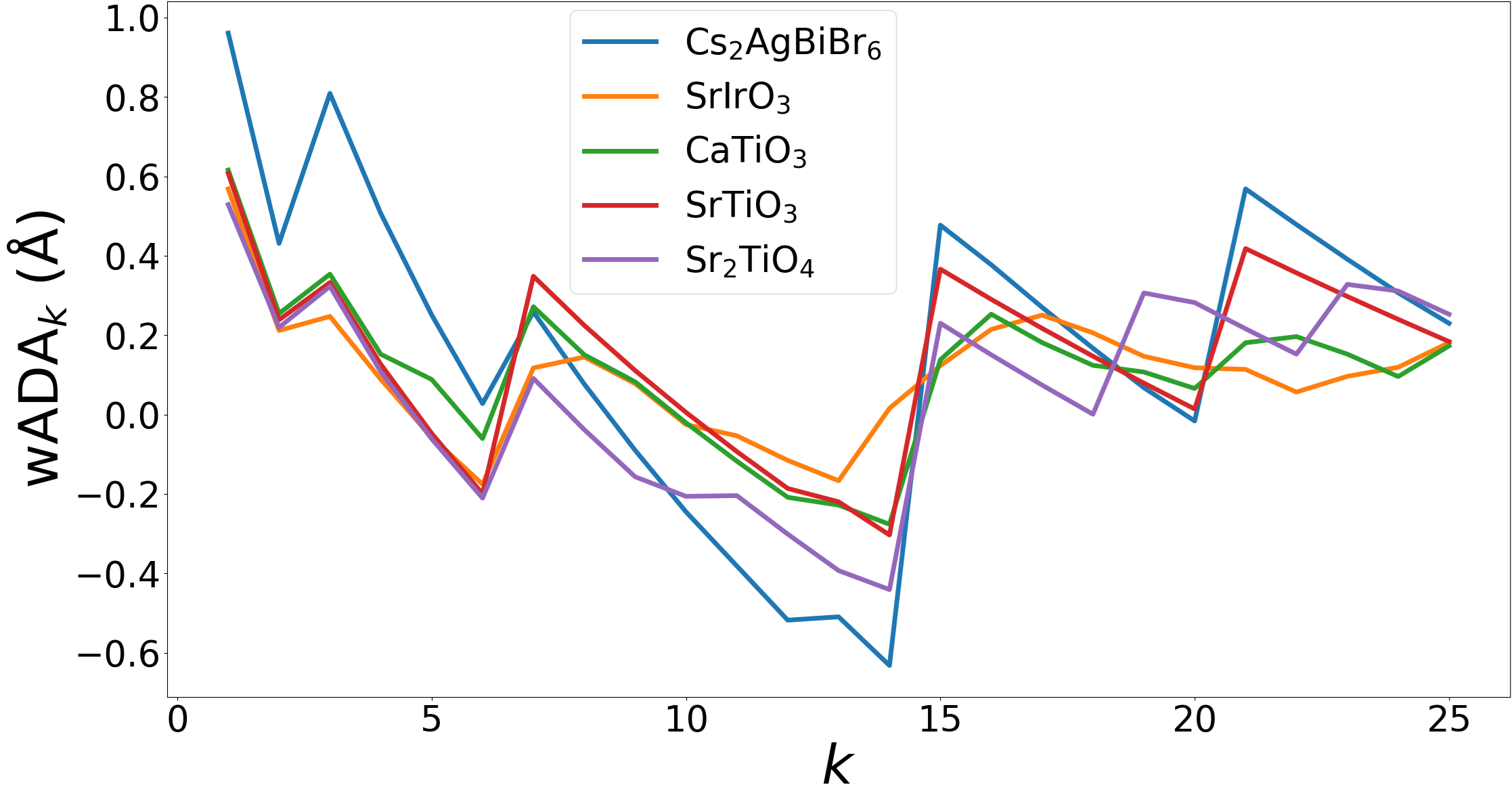}
\label{fig:AMD25ADA_5perovskites}
\end{figure}

\begin{figure}[h!]
\caption{Distances in Angstroms between 5 simple crystals and 5 perovskites from Table~\ref{tab:MPcrystals}. \textbf{Upper triangle}: EMDs on weighted w$\PDA(S;25)$. \textbf{Lower triangle}: EMDs on unweighted $\PDA(S;25)$.}
\includegraphics[width=\textwidth]{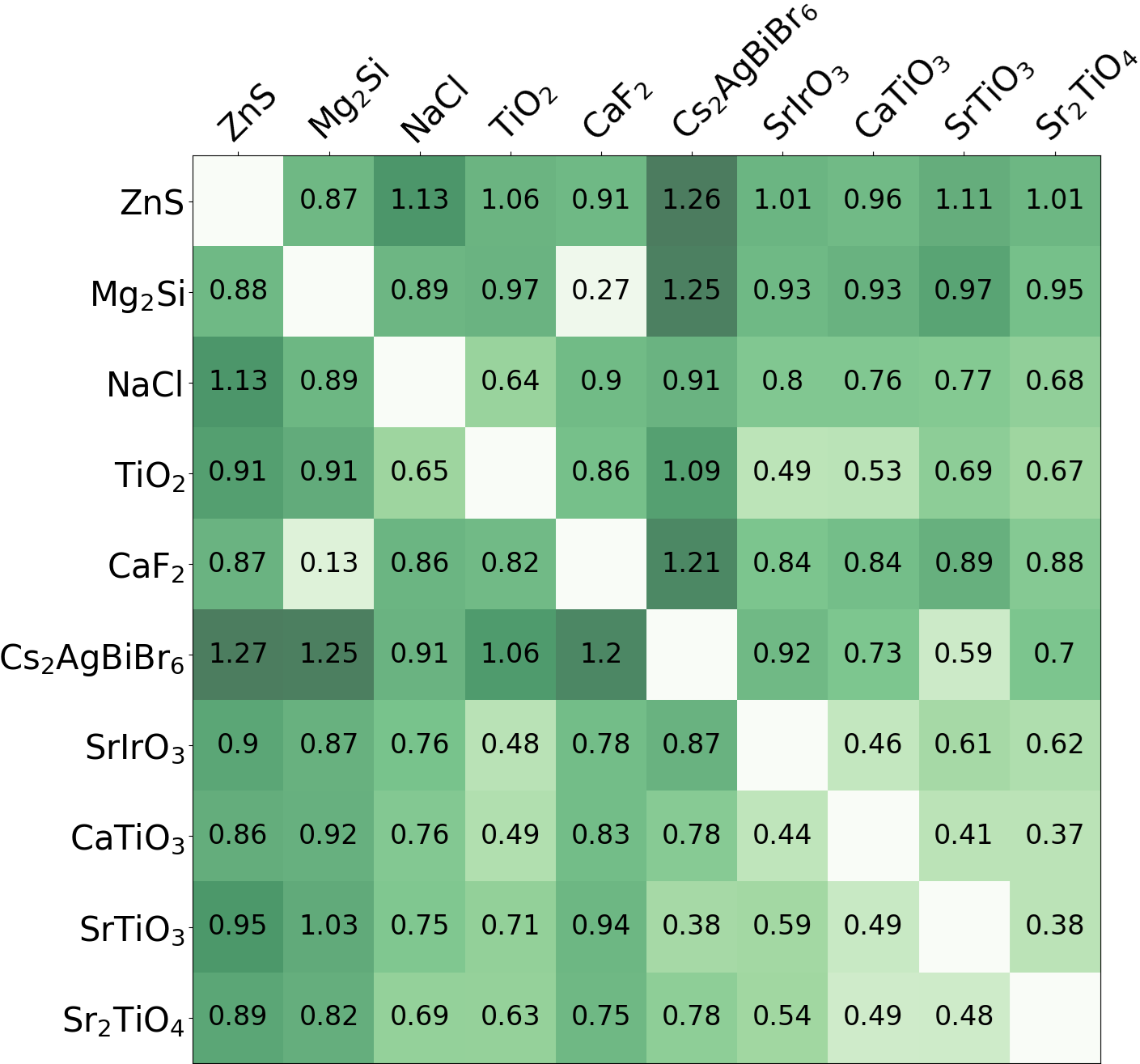}
\label{fig:EMD10crystals}
\end{figure}

{This appendix includes extra examples of invariant computations for 5 perovskites in addition to 5 simple crystals in Fig}.~\ref{fig:AMD25ADA_5common_crystals}, {see corresponding entries from the MP in Table}~\ref{tab:MPcrystals}, {instructions for running the Python code for all invariants, and high-resolution navigation maps.
The zip folder with supplementary information includes the Python code and tables with $\PDD$ and $\PDA$ matrices for the $5+5$ example crystals.}
Fig.~\ref{fig:AMD25ADA_5perovskites} and Table~\ref{fig:EMD10crystals} {compare unweighted (based on atomic centers) and weighted versions (including atomic masses) of invariants.
The atomic masses generally increase $\EMD$ distances but the geometry of atomic centers already captures chemistry.}   
\smallskip

\begin{table}[!h]
\begin{tabular}{l|lllll}
simple crystals & rock salt & rutile & zincblende & fluorite & antifluorite \\
composition &  \ce{NaCl} & \ce{TiO2} & \ce{ZnS} & \ce{CaF2} & \ce{Mg2Si} \\
MP entry id & 22862 & 2657 & 10695  & 2741 & 1367 \\
\hline
perovskites & cubic & hexagonal & tetragonal & double & Ruddlesden-Popper \\
composition &  \ce{SrTiO3} &  \ce{SrIrO3} &\ce{CaTiO3} & \ce{Cs2AgBiBr6} & \ce{Sr2TiO4} \\
MP entry id & 5229 & 17097 & 3442  & 1078250 & 5532 
\end{tabular}
\caption{{Names, compositions and IDs of $5+5$ crystals whose invariants are in Fig.}~\ref{fig:AMD25ADA_5common_crystals} and \ref{fig:AMD25ADA_5perovskites}.}
\label{tab:MPcrystals}
\end{table}

The instructions below will reproduce the distances between the structures in Fig.~\ref{fig:MnAgO2}, whose CIFs are publicly available at these URLs:
\begin{itemize}
    \item icsd\_670065.cif: \url{https://www.ccdc.cam.ac.uk/structures/Search?Ccdcid=670065&DatabaseToSearch=Published}
    \item icsd\_139006.cif: \url{https://www.ccdc.cam.ac.uk/structures/Search?Ccdcid=139006&DatabaseToSearch=Published}
    \item MnAgO2.cif: Under \verb|Supplementary-Data\Structure_Files| in the supplementary data at \url{https://www.nature.com/articles/s41586-023-06734-w#Sec14}
\end{itemize}

The supplementary materials also include the script \verb|compare_MnAgO2.py| for the distances. With Python 3.9+, install average-minimum-distance (1.5.3) \cite{AMD_DW} with
\begin{center}
    \verb| pip install average-minimum-distance==1.5.3|
\end{center}
Run the script with
\begin{center}
    \verb| python compare_MnAgO2.py|
\end{center}
The expected output is:
\begin{center}
\verb|>> EMD(PDA(MnAgO2, 100), PDA(icsd_670065, 100)) = 0.0975A| \\
\verb|>> EMD(PDA(MnAgO2, 100), PDA(icsd_139006, 100)) = 0.3675A|
\end{center}
\smallskip

The GNoME paper used the Pymatgen structure matcher \cite{pymatgen}, which cannot filter out near-duplicate structures according to the quoted steps below.
\smallskip

\noindent
``1. Given two structures: s1 and s2

\noindent
2. Optional: Reduce to primitive cells.

\noindent
3. If the numbers of sites do not match, return False.''
\smallskip

\noindent
These steps are followed by several heuristic steps which involve finding deviations between atoms in the reduced unit cell. If step 2 above is optionally missed, step 3 can output False (no match) for identical crystals given with different non-primitive cells.
If step 2 is enforced, step 3 will output False (no match) for any nearly identical crystals, whose primitive cells can arbitrarily differ due to a tiny atomic displacement as in Fig.~\ref{fig:perturbations}. 
For the above reasons, our findings show that this method of comparing structures was insufficient to filter out existing duplicates from {materials databases}, resulting in the AI silently reproducing {near-duplicates from} the training set.
\smallskip

{Definition}~\ref{dfn:EMD} {can use arbitrary weights of points so that the total weight within a unit cell is 1 after normalization.
When all equal rows of $\PDD(S;k)$ are collapsed to a single row, the discontinuity in Fig.}~\ref{fig:perturbations} {is properly resolved. 
The unweighted version of $\EMD$ on $\PDA$s below uses zero-sized points at atomic centers with equal weights, which 
can be multiplied (before normalization) by atomic masses or 
ionic radii.}
\smallskip

\begin{dfn}[Earth Mover's Distance $\EMD$ {\cite{rubner2000earth}}]
\label{dfn:EMD}
Consider any matrix $\PDA(S;k)$ as a distribution of rows $R_i(S)$ with weights $w_i(S)$ for $i=1,\dots,m(S)$ such that $\sum\limits_{i=1}^m w_i=1$. 
The \emph{Earth Mover's Distance} 
$\EMD(\PDA(S;k),\PDA(Q;k))=\min\limits_{f_{ij}} \sum\limits_{i=1}^{m(S)}\sum\limits_{j=1}^{m(Q)} f_{ij} L_\infty(R_i(S),R_j(Q))$ is minimized for all real $f_{ij}\geq 0$ (called \emph{flows}) subject to the conditions $\sum\limits_{i=1}^{m(S)}f_{ij}\leq w_j(Q)$, $\sum\limits_{j=1}^{m(S)}f_{ij}\leq w_i(S)$, $ \sum\limits_{i=1}^{m(S)}\sum\limits_{j=1}^{m(Q)} f_{ij}=1$.
\end{dfn}
 
 The first condition $\sum\limits_{j=1}^{m(Q)} f_{ij}\leq w_i(S)$ means that not more than the weight $w_i(S)$ of the component $R_i(S)$ `flows' into all components $R_j(Q)$ via `flows' $f_{ij}$ for $j=1,\dots,m(Q)$. 
The second condition $\sum\limits_{i=1}^{m(S)} f_{ij}=w_j(Q)$ means that all `flows' $f_{ij}$ from $R_i(S)$ for $i=1,\dots,m(S)$ `flow' into $R_j(Q)$ up to the maximum weight $w_j(Q)$.
The last condition
$\sum\limits_{i=1}^{m(S)}\sum\limits_{j=1}^{m(Q)} f_{ij}=1$
 forces to `flow' all rows $R_i(S)$ to all rows $R_j(Q)$.  
\smallskip

\begin{proof}[Proof of Theorem~\ref{thm:LND}]
Let $S$ be obtained from a periodic point set $Q\subset \R^n$ by perturbing every point of $Q$ up to Euclidean distance $\ep$, which is smaller than a minimum half-distance between any points of $Q$.
Then $S,Q$ have a common lattice by \cite[Lemma~4.1]{edelsbrunner2021density} and hence the same number $m$ of points in a common unit cell, and equal Point Packing Coefficients $\PPC(S)=\PPC(Q)$ from Definition~\ref{dfn:PPC}.
\smallskip

Since Definition~\ref{dfn:PDA} uses the $L_\infty$ metric on rows of $\PDA$s, the Earth Mover's Distance is unaffected by subtracting the same term $\PPC\sqrt[3]{k}$, so $\EMD(\PDD(S;k),\PDD(Q;k))=\EMD(\PDA(S;k),\PDA(Q;k))$.
Then \cite[Theorem~4.3]{widdowson2022average} implies that
$\EMD(\PDA(S;k),\PDA(Q;k)|\leq 2\ep$.
The minimum for all sets $Q$ in a finite dataset $D$ can not be larger, so $\LND(S;D)\leq 2\ep$ by Definition~\ref{dfn:LND}.
\smallskip

Conversely, assume that $S$ is obtained from $Q\in D$ by perturbing every atom of $Q$ up to Euclidean distance $\ep<0.5\LND(S;D)<r(Q)$.
The previously proved inequality implies that $\LND(S;D)\leq 2\ep<\LND(S;D)$, which is a contradiction. 
\end{proof}

The figures below include high resolution images from Fig.~\ref{fig:A-lab_on_ICSD+MP}.

\vspace*{-2mm}

\begin{figure}[h!]
\caption{
A-lab crystals in cyan over the ICSD and MP heatmap in the coordinates (density, $\ADA_1$).
}
\includegraphics[width=\textwidth]{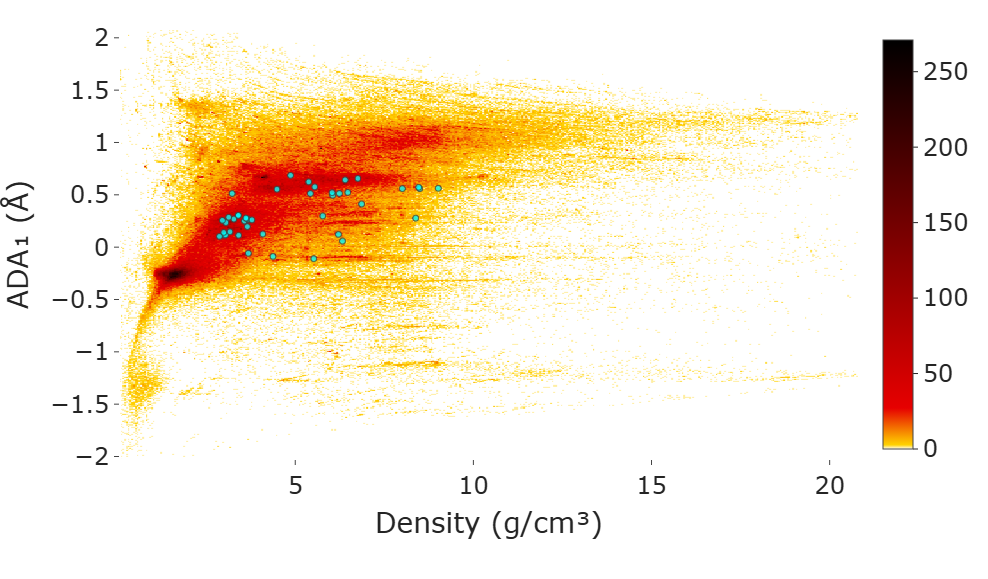}
\label{fig:density_ADA1}
\end{figure}

\begin{figure}[h!]
\caption{
A-lab crystals in cyan over the ICSD and MP heatmap in the coordinates ($\ADA_2$, $\ADA_3$).
}
\includegraphics[width=\textwidth]{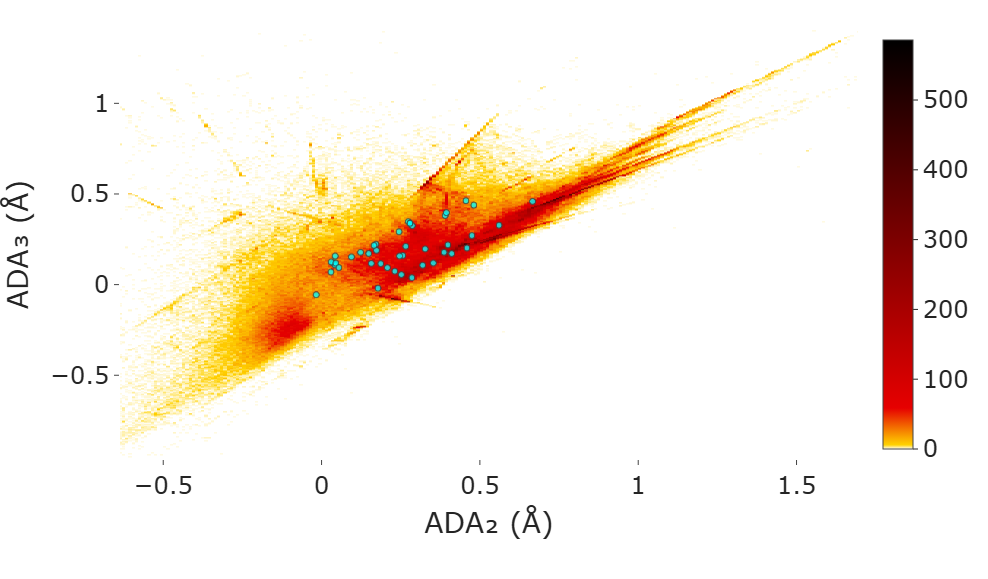}
\label{fig:ADA2_ADA3}
\end{figure}

\begin{figure}[h!]
\caption{
A-lab crystals in cyan over the ICSD and MP heatmap in the coordinates ($\ADA_4$, $\ADA_5$).
}
\includegraphics[width=\textwidth]{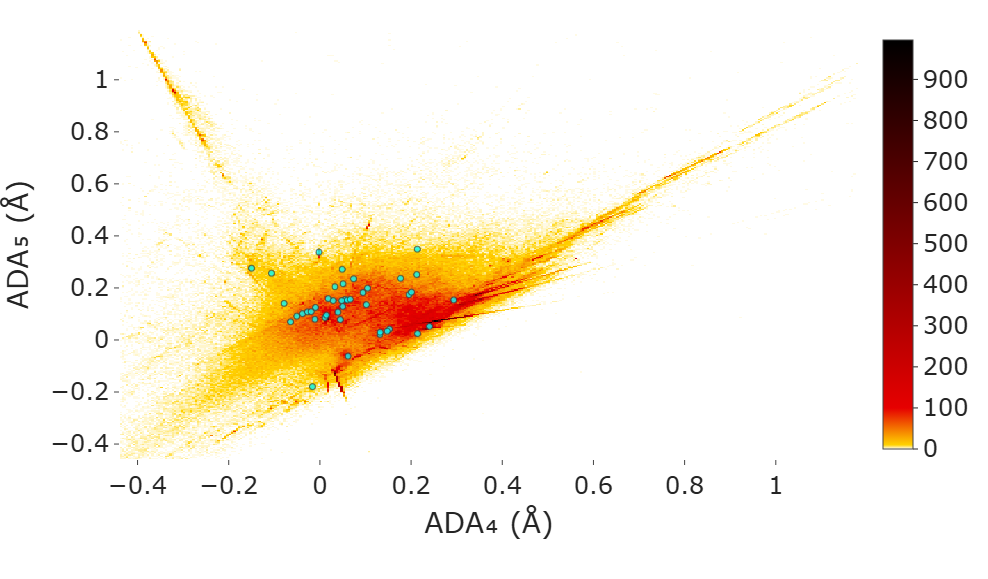}
\label{fig:ADA4_ADA5}
\end{figure}

\begin{figure}[h!]
\caption{
A-lab crystals in cyan over the ICSD and MP heatmap in the coordinates ($\ADA_{1}$, $\ADA_{20}$).
}
\includegraphics[width=\textwidth]{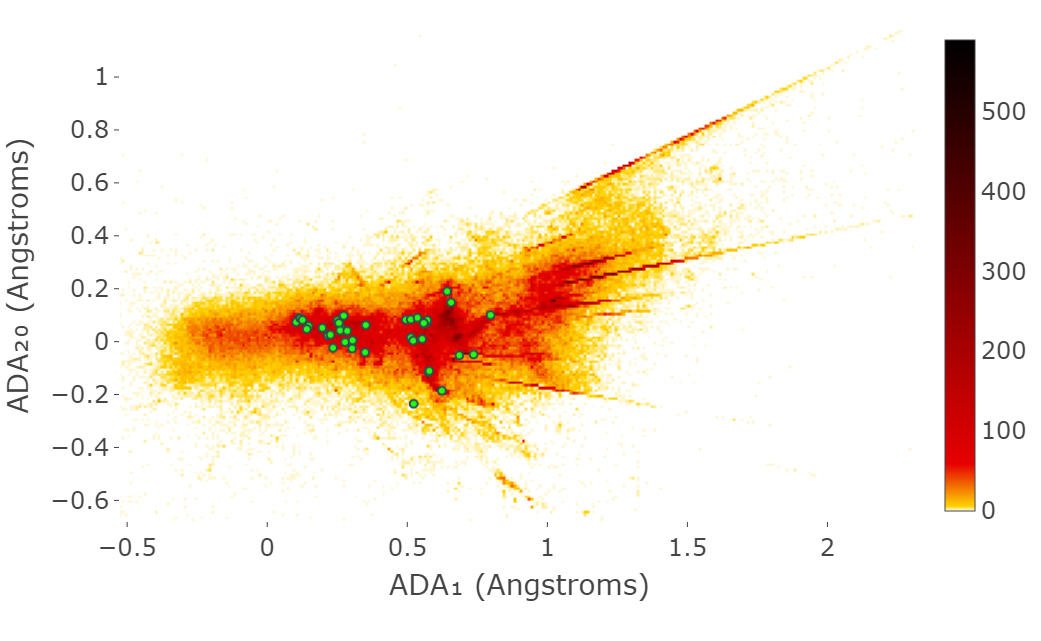}
\label{fig:ADA1_ADA20}
\end{figure}

\begin{figure}[h!]
\caption{
A-lab crystals over the ICSD and MP heatmap in the coordinates ($\ADA_{20}$, $\ADA_{100}$).
}
\includegraphics[width=\textwidth]{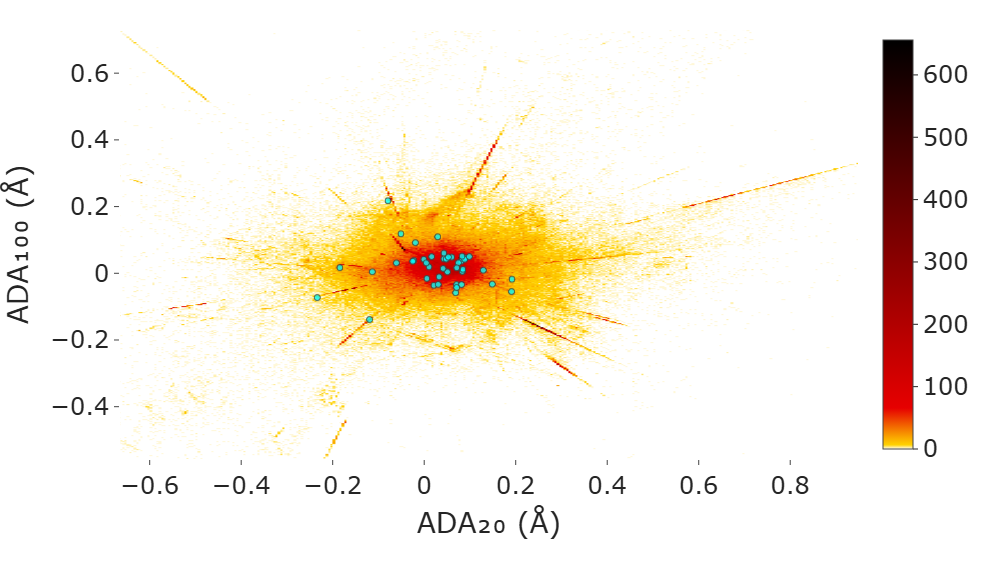}
\label{fig:ADA20_ADA100}
\end{figure}

\end{appendices}

\end{document}